\documentclass[longauth]{aa}
\usepackage{linenoaa}
\usepackage{rotating}
\usepackage{graphicx}
\usepackage{booktabs}
\usepackage{txfonts}
\usepackage{natbib} 
\bibpunct{(}{)}{;}{a}{}{,}
\usepackage{graphicx}
\usepackage{caption}
\usepackage{subcaption}
\usepackage{footnote}
\makesavenoteenv{table*}
\usepackage{threeparttable}
\usepackage{txfonts}
\usepackage{tikz}
\usepackage{amsmath}
\usepackage{soul}
\usepackage{hyperref}
\hypersetup{
    colorlinks=true,
    linkcolor=blue,
    citecolor=blue,
    urlcolor=blue
}

\begin{document}
\title{Mapping the obscured Galactic disk: ERIS J-band low-resolution spectroscopy of 27 Classical Cepheids}
\author{A.~Nunnari\inst{1}\fnmsep\inst{2}\fnmsep\inst{3}\and 
           C.~Lardo\inst{4} \and
           V.~D'Orazi\inst{2}\fnmsep\inst{1}\and
           V.~F.~Braga\inst{1}\and
           G.~Fiorentino\inst{1}\and
           G.~Bono\inst{2}\fnmsep\inst{1}\fnmsep\inst{5}\and
           M.~Fabrizio\inst{6}\fnmsep\inst{5}\and
           R.-P.~Kudritzki\inst{7}\fnmsep\inst{8}\and
           E.~Sextl\inst{8}\and
           M.~Schultheis\inst{9}\and
           N.~Alvarez-Baena\inst{2}\fnmsep\inst{1}\fnmsep\inst{3}\and
           K.~Baeza-Villagra\inst{2}\fnmsep\inst{1}\fnmsep\inst{3}\and    
           G.~B\"{o}cek~Topcu\inst{2}\fnmsep\inst{1}\and
           C.~Campa\inst{1}\fnmsep\inst{2}\fnmsep\inst{3}\and
           G.~Ceci\inst{1}\fnmsep\inst{2}\fnmsep\inst{3}\and
           R.~da~Silva\inst{1}\fnmsep\inst{6}\and
           M.~Dall'Ora\inst{10}\and
           M.~Di~Criscienzo\inst{1}\and
           C.~Gavetti\inst{11}\fnmsep\inst{1}\and
           M.~Gholami\inst{10}\and
           S.~Leccia\inst{10}\and
           A.~Liberatori\inst{12}\and
           M.~Marconi\inst{10}\and
           N.~Matsunaga\inst{13},  
           R.~Molinaro\inst{10}\and
           M.~Monelli\inst{14}\fnmsep\inst{15}\fnmsep\inst{16}\and
           A.~Mucciarelli\inst{4}\fnmsep\inst{17}\and
           I.~Musella\inst{1}\and
           J.~Olivares~Carvajal\inst{10}\and
           A.~M.~Piersimoni\inst{14},
           A.~Pietrinferni\inst{14},
           B.~Pozo~Vallejos\inst{2}\fnmsep\inst{1}\fnmsep\inst{3}\and
           V.~Ripepi\inst{10}\and
           M. Sànchez-Benavente\inst{15}\fnmsep\inst{16},
           M.~Tantalo\inst{1}\and           
           F.~Th\'ev\'enin\inst{9}\and
           E.~Valenti\inst{18}\fnmsep\inst{19}\and
           M.~Zoccali\inst{20}\and
           G.~Agapito\inst{21}\and
           M.~Black\inst{22}\and
           M.~Bonaglia\inst{21}\and
           R.~Briguglio\inst{21}\and
           Y.~Cao\inst{23}\and
           L.~Carbonaro\inst{21}\and
           G.~Cresci\inst{21}\and
           Y.~Dallilar\inst{24}\and
           R.~Davies\inst{23}\and
           M.~Deysenroth\inst{23}\and
           I.~Di~Antonio\inst{14}\and
           A.~Di~Cianno\inst{14}\and
           G.~Di~Rico\inst{14}\and
           M.~Dolci\inst{14}\and
           F.~Eisenhauer\inst{23}\and
           S.~Esposito\inst{21}\and
           D.~Ferruzzi\inst{21}\and
           H.~Feuchtgruber\inst{23}\and
           N.~F\"{o}rster-Schreiber\inst{23}\and
           R.~Genzel\inst{23}\and
           S.~Gillessen\inst{23}\and
           P.~Grani\inst{21}\and
           M.~Hartl\inst{23}\and
           K.~Kravchenko\inst{23}\and
           D.~Lutz\inst{23}\and
           M.~Macintosh\inst{22}\and
           T.~Ott\inst{23}\and
           D.~Pearson\inst{22}\and
           A.~Puglisi\inst{21}\and
           S.~Rabien\inst{23}\and
           C.~Rau\inst{23}\and
           A.~Riccardi\inst{21}\and
           T.~Shimizu\inst{23}\and
           E.~Sturm\inst{23}\and
           L.~Tacconi\inst{23}\and
           M.~Xompero\inst{21}
           }
   \institute{
          INAF -- Osservatorio Astronomico di Roma, Via Frascati 33, 00078 Monte Porzio Catone, Italy.
          \and 
          Dipartimento di Fisica, Università di Roma Tor Vergata, Via della Ricerca Scientifica 1, 00133 Rome, Italy 
          \and
          Dipartimento di Fisica, Sapienza Università di Roma, P.le A. Moro 5, Roma 00185, Italy
          \and
          Dipartimento di Fisica e Astronomia, Università degli Studi di Bologna, Via Gobetti 93/2, 40129 Bologna, Italy
          \and
          INFN Sezione di Roma2, Università di Roma Tor Vergata, Via della Ricerca Scientifica, 1 , Roma, 00133, Italy
          \and
          ASI -- Space Science Data Center, via del Politecnico snc, I-00133 Roma, Italy
          \and 
          Institute for Astronomy, University of Hawai’i at Manoa, Honolulu, HI 96822, USA
          \and
          LMU M\"{u}nchen, Universit\"{a}tssternwarte, Scheinerstr. 1, 81679 M\"{u}nchen, Germany
          \and
          Universit\'e Côte d’Azur, Observatoire de la Côte d’Azur, CNRS, Laboratoire Lagrange, France
          \and 
          INAF -- Osservatorio Astronomico di Capodimonte, salita Moiariello 16, 80131, Napoli, Italy
          \and 
          Dipartimento di Matematica e Fisica, Università degli Studi Roma Tre, Via della Vasca Navale 84, 00100 Roma, Italy
          \and
          National and Kapodistrian University of Athens, Panepistimiopolis, 15784 Zografos, Athens, Greece
          \and
          Department of Astronomy, School of Science, The University of Tokyo, 7-3-1 Hongo, Bunkyo-ku, Tokyo 113-0033, Japan
          \and
          INAF -- Osservatorio Astronomico di Abruzzo, Via M. Maggini, 64100 Teramo, Italy
          \and
          IAC -- Instituto de Astrof\'isica de Canarias, Calle V\'ia Lactea s/n, E-38205 La Laguna, Tenerife, Spain
          \and
          Departmento de Astrof\'isica, Universidad de La Laguna, E-38206 La Laguna, Tenerife, Spain
          \and
          INAF -- Osservatorio di Astrofisica e Scienza dello Spazio di Bologna, INAF, Via Gobetti 93/3, Bologna, I-40129, Italy
          \and
          European Southern Observatory, Karl-Schwarzschild-Strasse 2, 85748 Garching bei Munchen, Germany
          \and
          Excellence Cluster ORIGINS, Boltzmann–Strasse 2, D–85748 Garching bei München, Germany
          \and
          Pontificia Universidad Cat\'olica de Chile, Instituto de Astrof\'isica, Av. Vicuña Mackenna 4860, 782-0436 Macul, Santiago, Chile
          \and
          INAF -- Osservatorio Astrofisico di Arcetri, Largo E. Fermi 5, 50125, Firenze, Italy
          \and
          STFC UK ATC, Royal Observatory Edinburgh, Blackford Hill. Edinburgh, EH9 3HJ, UK
          \and
          Max Planck Institute for extraterrestrial Physics, Giessenbachstraße 1, 85748 Garching, Germany
          \and
          I. Physikalisches Institut, Universit\"{a}t zu K\"{o}ln, Z\"{u}lpicher Str. 77, 50937, K\"{o}ln, Germany
   }
   \date{Received Month DD, YYYY}

  \abstract{Classical Cepheids (CCs) are standard candles and key tracers of young stellar populations. However, their use in the most obscured regions is hampered by the high interstellar extinction at optical wavelengths. Near-infrared spectroscopy offers major advantages by greatly reducing absorption corrections and enabling access to regions of the Milky Way that are still poorly sampled. In this work, we determine the atmospheric parameters and abundances of Fe, Mg and Si for a sample of 27 CCs based on observations with ERIS@VLT in J-band low-resolution mode (R$\sim$4000). Combining PySME with 1D MARCS atmosphere models and non local thermal equilibrium grids, we simultaneous optimise the atmospheric parameters through spectral fitting. This method is validated with a direct comparison with high-resolution optical investigations and re-analysis of archive optical spectra for the targets in common. The results show good consistency in both metallicity and chemical abundances. We obtain mean abundance ratios of $\mathrm{\langle[Mg/Fe]\rangle=0.01\pm0.12}$~dex and $\mathrm{\langle[Si/Fe]\rangle=0.10\pm0.09}$~dex, in agreement with reference samples. Moreover, we provide for the first time the Mg and Si abundances of 12 CCs. Finally, we have found distinct kinematic properties for two of the most distant Galactic CCs, located at Galactocentric distances of $\approx$~28--29~kpc and with extreme values of vertical height, $\approx$+4.8~kpc and $\approx-$5.3~kpc; more data are required to constrain their nature. Our results confirm that J-band spectroscopy at low-resolution can be used as a good diagnostic for the chemical investigation in highly obscured Galactic regions, otherwise inaccessible at optical wavelengths.}
\authorrunning{A. Nunnari et al.}
   \maketitle
\nolinenumbers

\section{Introduction}
Classical Cepheids (CCs) are cornerstone distance indicators and powerful tracers of the structure and recent chemical evolution of the Milky Way (MW) and nearby galaxies. These radially pulsating variables are intermediate-mass stars (approximately 4 to 11~$M_{\odot}$) in the central helium-burning phase, known as the "blue loop" (see \citealt{bhardwaj20} and \citealt{bono2024cepheids} for recent reviews). CCs provide a unique high-resolution map of the chemical enrichment history of the Galactic thin disk due to their youth (age~<~300~Myr), their wide spatial distribution (from $\sim$~4 up to $\sim$~30~kpc, \citealt{da2023oxygen}; \citealt{nunnari26}; \citealt{trentin26}), and their well-known distances thanks to the period-luminosity relation (\citealt{freedman24}) and Gaia parallaxes (\citealt{ripepi22, ripepi26, GaiaCollaboration2023DR3}).

Although most chemical studies of Cepheids have been conducted in the optical range (see e.g. \citealt{genovali15, luck18, kovtyukh22, da2023oxygen, Trentin24B, trentin26, nunnari26}), near-infrared (NIR) spectroscopy offers decisive advantages for probing the structure of the Galaxy. The main benefit is the drastic reduction in interstellar extinction at longer wavelengths, allowing for observation of Cepheids in highly obscured regions, such as the inner Galactic disk (Galactocentric radius, $\mathrm{R_{GC}}$~<~5~kpc) or the far side of the Galactic plane ($\mathrm{R_{GC}}$~>~20~kpc), which are hard to reach with optical instruments (\citealt{ryde07}; \citealt{matsunaga23}; \citealt{catanzaro26}). As a result, the radial chemical gradients constrained by the CCs suffer from limited sampling in the innermost and outermost regions of the disk. Furthermore, NIR period-luminosity relations are intrinsically more accurate than optical ones because they are less sensitive to the temperature width of the instability strip and to variations in bolometric corrections (\citealt{bono2024cepheids}).

However, spectroscopic abundance analysis in NIR bands presents significant technical challenges. Compared to the optical coverage, the number of available atomic lines is considerably limited (\citealt{proxauf2018new}; \citealt{senturk26}). In particular, in the J-band the overall scarcity of spectral features makes it difficult to simultaneously and accurately constrain the main atmospheric parameters (\citealt{bergemann12a}; \citealt{gazak14}; \citealt{lardo15}; \citealt{davies17}), i.e., effective temperature ($\mathrm{T_{eff}}$), surface gravity ($\mathrm{\log g}$), metallicity ($\mathrm{[M/H]}$) and micro-turbulence ($\mathrm{v_{mic}}$). However, this spectral domain offers well-isolated iron and $\alpha$ elements (Mg, Si, Ca, Ti) lines. Many published NIR line lists, including the J-band lines, suffer from uncertainties in oscillator strengths  ($\mathrm{log} \,gf$) and lack laboratory validation as extensive as that available for optical transitions (\citealt{andreasen16, kondo19, ryde19, elgueta24, senturk26}). This issue is particularly evident in low-resolution spectra, where the blending of the few available features further complicates the identification of discrete spectral lines (\citealt{davies10}; \citealt{gazak14}). Nevertheless, different studies highlighted the potential of low-resolution J-band spectroscopy (e.g. \citealt{bergemann12a, bergemann13, bergemann15}; \citealt{gazak14, gazak15}; \citealt{patrick15, patrick16, patrick17}), especially for estimating the atmospheric parameters in red supergiants.

In recent years, several studies have begun to explore the potential of NIR spectroscopy for Cepheids. The calibration of these methods for Cepheids is made even more complex by their intrinsic variability. During a single pulsation cycle, a Cepheid can experience temperature variations of up to 1000~K and changes in surface gravity of nearly 1.0~dex (\citealt{proxauf2018new}; \citealt{da2022new}). \citet{inno19} provided the first metallicity measurements for obscured Cepheids in the inner disk using medium-resolution spectra (R~$\sim$~3100) obtained with ISAAC@VLT. \citet{minniti20} used NIR spectra of 30 CCs from X-SHOOTER@VLT (R~$\sim$~8000) to derive the radial metallicity gradient at the far side of the disk; a work carried out in \citet{minniti21} where the authors traced the spiral structure. 
\citet{matsunaga23} extended the analysis to 16 CCs in the inner disk ($\mathrm{R_{GC}}$~=~3--5.6~kpc) using high-resolution spectra (R~$\sim$~28000) in the Y-J bands obtained with WINERED@Magellan-Clay, thereby determining the metallicity gradient in this critical region for the first time. \citet{elgueta24} provided an astrophysical calibration of the $\mathrm{\log}\,gf$ values for spectral lines in Y-J band based on the optical  abundances reported by \citet{luck18}, laying the groundwork for a consistent metallicity scale between optical and infrared spectroscopy. This work has been extended in \citet{elgueta26}, where the authors provided a robust infrared line list for abundance determination, based on Gaia FGK benchmark stars. Recently, \citet{catanzaro26} validated a list of NIR lines for abundance analysis in H-K bands, although their atmospheric parameters were derived in part through photometric approaches or empirical calibrations. They used spectra of 23 CCs from IGRINS@Gemini with R~$\sim$~45000. Finally, \citet{urbaneja26B} validated Cepheid metallicities against the optical reference scale using low-resolution SpeX/IRTF (Y-J band, R~$\sim$~2000) spectra. The spectra were analysed with a Bayesian full-spectrum method based on MARCS model atmospheres (\citealt{gustafsson2008grid}), TURBOSPECTRUM spectral synthesis (\citealt{alvarez98, plez12}), and the MAUI framework (\citealt{urbaneja26}). 

In this study, as part of the "RR Lyrae In and Beyond the Galactic Bulge survey" (RIBBS) program, we present a spectral synthesis method to simultaneously derive atmospheric parameters and Mg and Si abundances for 27 CCs discovered by OGLE-IV (\citealt{udalski15}). Using low-resolution ERIS@VLT J-band spectra (R~$\sim$~4000), we demonstrate that reliable parameter optimization along with Mg and Si abundances is achievable even in this challenging regime. This work focuses on validating this approach by comparing it with high-resolution optical studies. In §~\ref{sec:datasetmethod} we present the sample along with the distance estimation, and the spectral synthesis method, focusing in particular on the calibration and validation of our approach. In §~\ref{sec:results} we discuss the main results of this study, starting from the kinematics and moving toward the chemical properties. Finally, in §~\ref{sec:summary} we summarise the findings of this study and propose future directions.


\section{Dataset and Method}
\label{sec:datasetmethod}
\subsection{Observations and Data Reduction}
As part of the RIBBS program, we observed 27 CCs using SPIFFIER, the integral field spectrograph of ERIS (\citealt{davies23}) mounted at the VLT, operating in the low-resolution J-band (R~$\sim$~4000) and LGS mode. Our targets span magnitudes of J=6.5-12.5~mag. For reaching a signal-to-noise ratio (SNR) of $\approx$~100, we adopted an exposure time of $\approx$~30 to $\approx$~900 seconds for the brightest and faintest targets, respectively.

\begin{figure*}
        \centering
    \includegraphics[width=1.\linewidth]{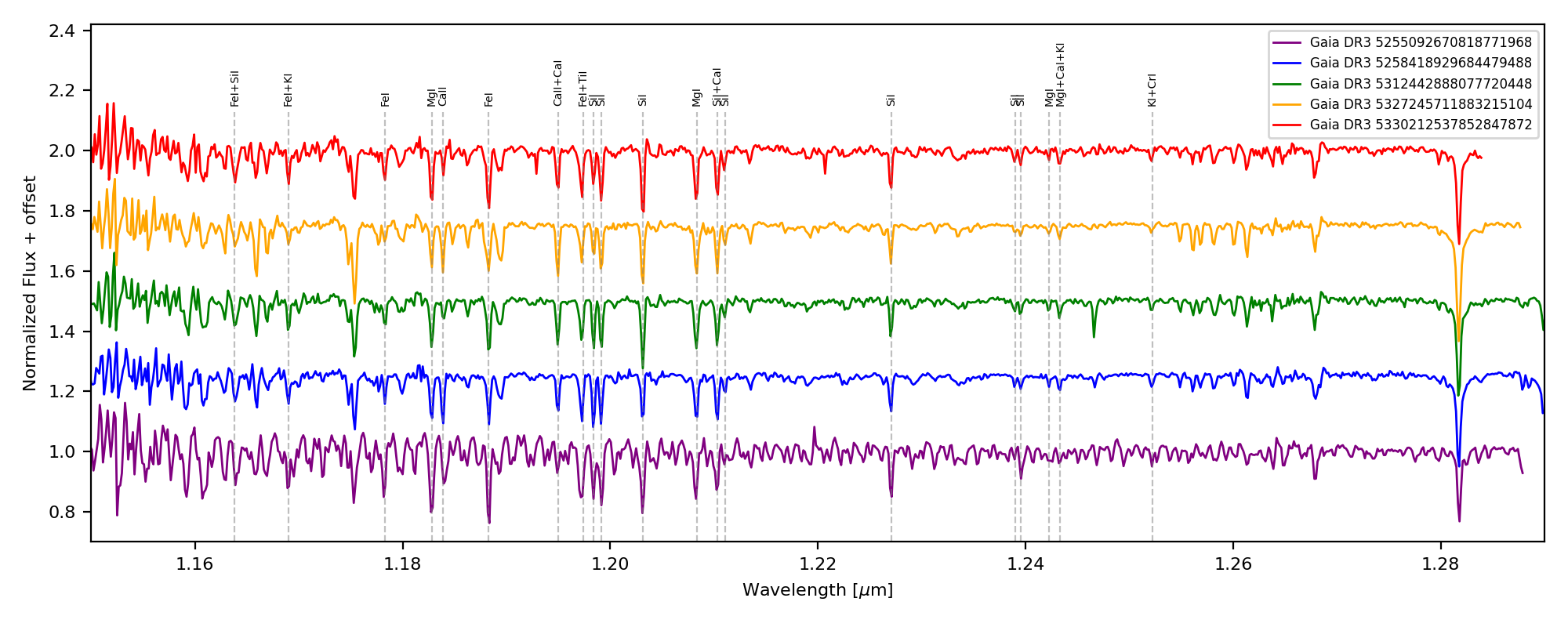}
    \caption{\small{Example of five ERIS J-band spectra of the current sample. The marked lines are those employed in the spectral fitting.}}
    \label{fig:spectra_plot}
\end{figure*}

To maximise the SNR, we extracted the spectra from the 15 brightest spaxels for each target, combining them via a median to produce a 1D spectrum. Using a larger number of spaxels does not result in a significant improvement in the SNR of the median spectrum. To remove atmospheric absorption features while preserving the integrity of the stellar lines, telluric correction was performed using synthetic telluric spectra from TAPAS\footnote{\url{https://tapas.aeris-data.fr/en/home/}}. The continuum normalization and radial velocity correction, along with the removal of telluric lines, have been performed using \texttt{iSpec} software\footnote{\url{https://www.blancocuaresma.com/s/iSpec}}(\citealt{blancocuaresma14}).  An example of five 1D spectra of our sample is shown in Fig.~\ref{fig:spectra_plot}. 

\subsection{Atmospheric Parameter and Abundance Estimation}

The spectroscopic analysis was performed using \texttt{PySME} (version 1.0.1, \citealt{wehrhahn21, jian26}), the Python-based implementation of the "Spectroscopy Made Easy" code (\citealt{piskunov17}). This tool determines optimal stellar parameters by fitting synthetic spectra to observed data through a $\chi^2$ minimization process. PySME provides the uncertainties according to the least-squares fit. We used a grid of MARCS 1D hydrostatic model atmospheres (\citealt{gustafsson2008grid}) in spherical geometry, computed in local thermodynamic equilibrium (LTE). We incorporated Non-LTE (NLTE) departure coefficients from the state-of-the-art grids for H, Li, C, N, O, Na, Mg, Al, Si, K, Ca, Ti, Mn, Fe, and Cu, provided by \citet{amarsi2020galah, amarsi22, amarsi25}, \citet{mallinson24}, and \citet{caliskan25}, since systematic departures from LTE are significant in supergiant atmospheres (\citealt{bergemann25}). The use of a NLTE approach is essential for the spectral analysis of CCs, since these young supergiant stars are subject to significant systematic distortions when using standard LTE models. These deviations affect the determination of fundamental atmospheric parameters, especially $\mathrm{v_{mic}}$ (\citealt{nunnari26}).

While our general approach follows \citet{nunnari26}, we implemented a specialised optimization strategy to handle the limited number of lines in the J-band. In a two-step procedure we optimize $\mathrm{T_{eff}}$, $\mathrm{\log(g)}$, $\mathrm{[M/H]}$\footnote{[M/H] is the global metallicity parameter, whose value is used to scale the individual chemical abundances.} and $\mathrm{v_{mic}}$:
\begin{enumerate}
    \item A preliminary simultaneous fit of $\mathrm{T_{eff}}$, $\mathrm{\log(g)}$, $\mathrm{[M/H]}$ and $\mathrm{v_{mic}}$;
    \item A refined optimization where the surface gravity was constrained by using a semi-empirical relation with $\mathrm{T_{eff}}$ and pulsation period provided by \citet{elgueta24}, following \citet{Groenewegen20} and \citet{matsunaga23}:
    \begin{equation*}
        \mathrm{\log g_0 = 6.483\cdot \log (T_{eff}/5800) - 0.775 \cdot \log P + 2.475}, 
    \end{equation*}
    which has a scatter of 0.108 dex.
\end{enumerate}
We fitted the parameters in the following ranges: $\mathrm{T_{eff}}$ (4750, 7000 K); $\mathrm{\log g}$ ($\mathrm{\log g_0-}$0.5, $\mathrm{\log g_0+}$0.5 dex); [M/H] ($-$1, 1 dex); $\mathrm{v_{mic}}$ (1, 7~$\mathrm{km~s^{-1}}$), using as starting values $\mathrm{T_{eff}}=5500$~K, $\mathrm{\log g}=1.5$~dex, $\mathrm{[M/H]}=0$~dex and $\mathrm{v_{mic}=3.5~km~s^{-1}}$, which are typical values for Galactic CCs (\citealt{markova25}). We used as initial chemical abundances solar values from \citet{asplund09}. We updated C and N initial abundances in order to have super-solar [N/Fe] and subsolar [C/Fe], typical of stars that have
experienced mixing episodes during their evolution such as
CCs, we used A(N)~$=8.23$ and A(C)~$=8.20$. 

Once the atmospheric parameters were optimised, we used the same line list to measure the individual abundances. In PySME we used the individual abundances as fitting parameters, obtaining the final iron ratio ([Fe/H]), magnesium ratio ([Mg/Fe]), and silicon ratio ([Si/Fe]). Further details on the spectral analysis method are provided in \citet{nunnari26}.

\subsection{Calibration and Validation}

The accuracy of our parameters depends on the quality of the line list employed, including the atomic data information. We  provide the wavelength in $\AA$, the oscillator strength ($\mathrm{\log\,} gf$) and the excitational potential ($\mathrm{E_{low}}$) in eV for each line in Table \ref{tab:linelist}. This J-band line list focuses on isolated atomic lines of iron and $\alpha$-elements (Mg, Si), while any molecular features are too weak. We used the VALD (\citealt{ryabchikova15}) database for atomic parameters, which includes a comprehensive collection of atomic and molecular transition parameters of astronomical interest. At a resolution of R~$\sim$~4000 ($\Delta \lambda \sim$~0.3~nm), many spectral features are subject to blending. Our blending investigation (Appendix~\ref{App:blending}) quantified these effects by comparing synthesised individual line components with the observed blended features. 

\begin{figure}[h]
    \centering
    \includegraphics[width=1.\linewidth]{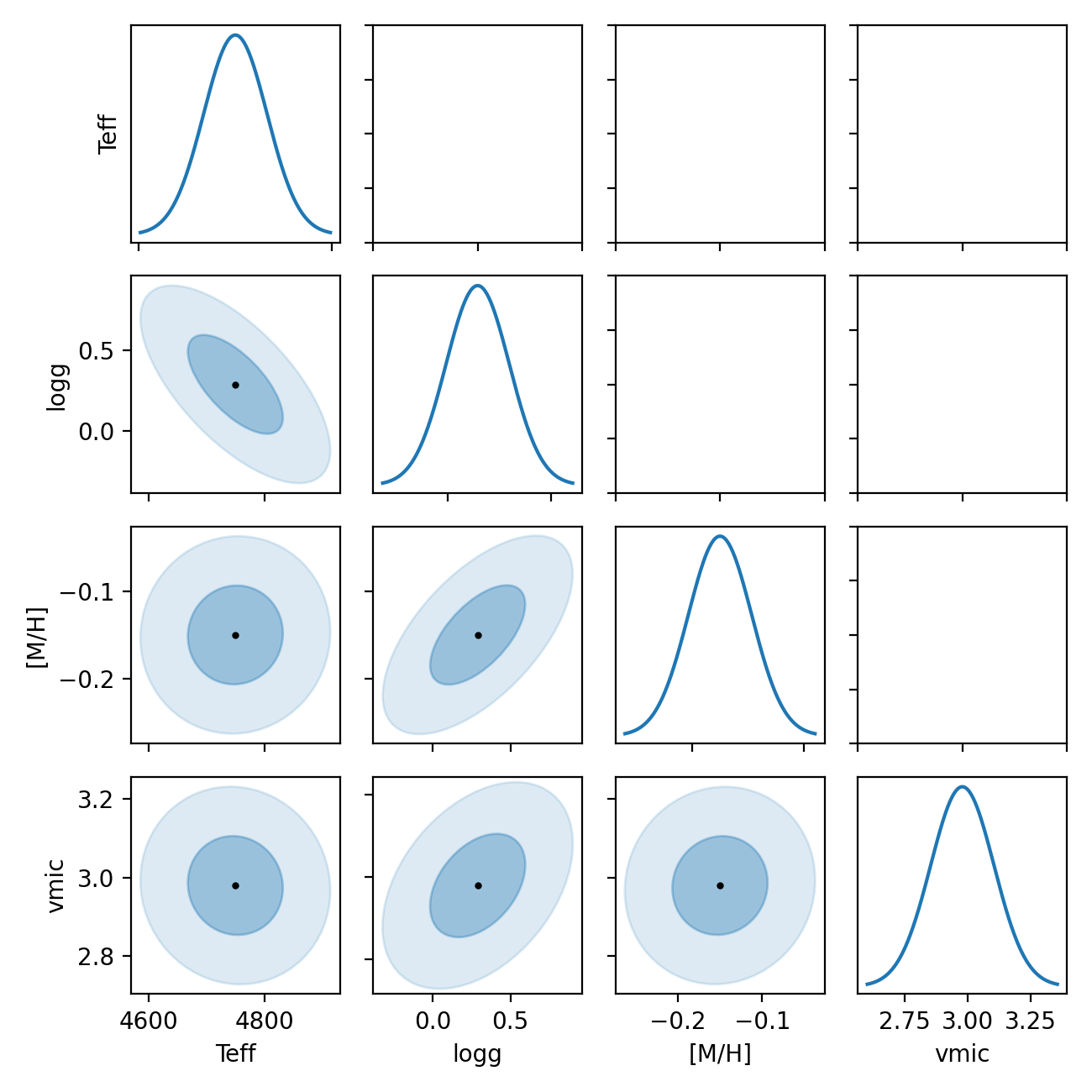}
    \caption{\small{Corner plot of the fitted parameters ($\mathrm{T_{eff}}$ in $\mathrm{K}$, $\mathrm{\log g}$ in dex, [M/H] in dex and $\mathrm{v_{mic}}$ in $\mathrm{km~s^{-1}}$) for Gaia DR3 ID 5255092670818771968. Contours at the 1 and 2$\sigma$ levels are shown.}}
    \label{fig:covar}
\end{figure}

\begin{table}[h]
    \centering
        \caption{\small{Line list employed for the atmospheric parameters and abundances estimation.}}
    \begin{tabular}{cccc}
    \hline
    \hline
       Line & Wavelength [$\AA$] & $\mathrm{\log }\,gf$ & $\mathrm{E_{low}}$ [eV] \\
       \hline
       Fe I & 11638.26 & $-$2.214 & 2.176 \\
       Fe I & 11689.97 & $-$2.068 & 2.223 \\
       Fe I & 11783.26 & $-$1.574 & 2.832 \\
       Mg I & 11828.17 & $-$0.333 & 4.346 \\
       Fe I & 11882.84 & $-$1.668 & 2.198 \\
       Si I & 11984.20 & 0.239 & 4.930 \\
       Si I & 11991.57 & $-$0.109 & 4.920 \\
       Si I & 12031.50 & 0.477 & 4.954 \\
       Si I & 12103.53 & $-$0.350 & 4.930 \\
       Si I & 12110.66 & $-$0.136 & 6.616 \\
       Si I & 12270.69 & $-$0.396 & 4.954 \\
       Mg I & 12433.45 & $-$0.967 & 5.933 \\
    \hline
    \hline
    \end{tabular}
    \label{tab:linelist}
    \tablefoot{Atomic species, central wavelength in $\AA$, oscillator strength ($\mathrm{\log}\,gf$) and excitation potential ($\mathrm{E_{low}}$) in eV.}
\end{table}
PySME returns the covariance matrix $C$ of the fitted parameters ($\mathrm{T_{eff}}$, $\mathrm{\log g}$, [M/H], $\mathrm{v_{mic}}$), estimated from the local curvature of the $\chi^2$ surface at the best-fit solution. The uncertainty on parameter $j$ is:
$$ \sigma_j = s\sqrt{C_{jj}}~,\quad s=\sqrt{\chi^2_\nu}$$

where $\chi^2_\nu$ is the reduced chi-square of the final fit, and $C_{jj}$ are the diagonal elements of the covariance matrix.\\
Since $C_{jj}$ is the marginalised variance, the correlations among parameters are already accounted for in $\sigma_j$. The flux uncertainties enter the $\chi^2$ as weights, so the S/N of the spectra is reflected in the coviarance matrix, and the rescaling by $s$ accounts for residual model mismatch. These are formal $1\sigma$ errors under a linear, Gaussian approximation; they do not include systematic effects (e.g. atomic parameters, model atmospheres, continuum placement). Fig.~\ref{fig:covar} shows the corner plot for Gaia DR3 5255092670818771968 with contours at the 1 and 2$\sigma$ levels. The strongest degeneracy involves $\mathrm{\log g}$, which is anticorrelated with $\mathrm{T_{eff}}$ ($\rho=-0.67$) and positively correlated with [M/H] ($\rho=0.61$) and $\mathrm{v_{mic}}$ ($\rho=0.40$). Variations in $\mathrm{T_{eff}}$, [M/H] or $\mathrm{v_{mic}}$ can be partly compensated by a change in $\mathrm{\log g}$ without significantly worsening the fit. The remaining pairs are nearly uncorrelated.

\begin{figure}
    \centering
    \includegraphics[width=1\linewidth]{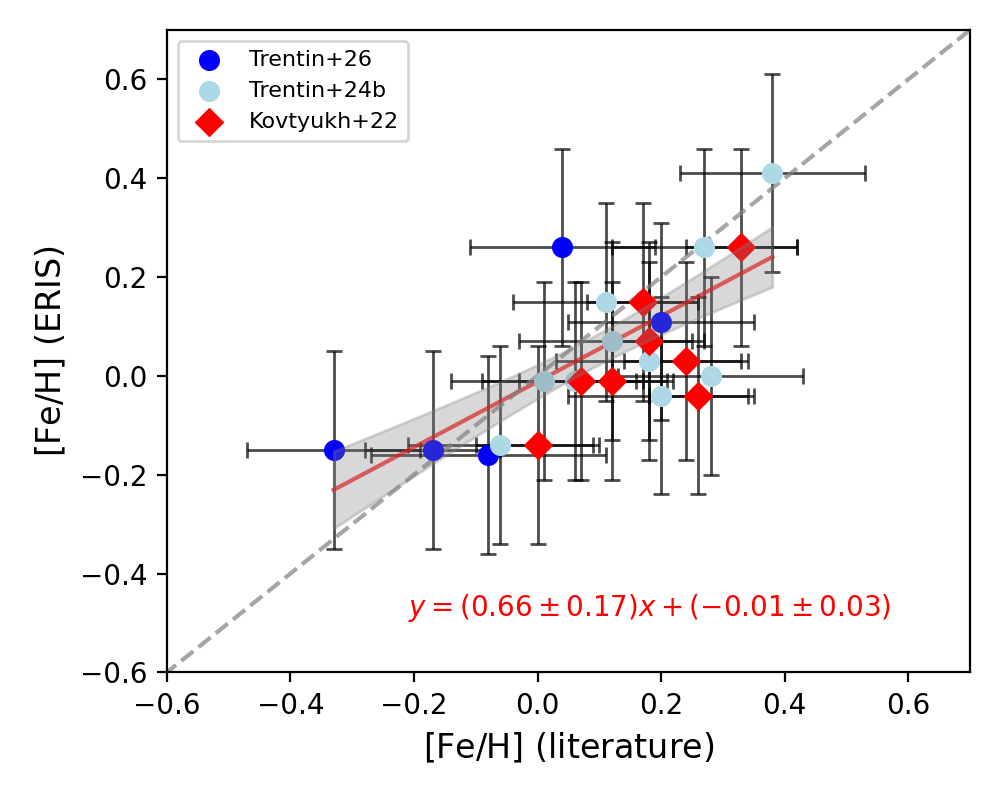}
    \caption{\small{Comparison of the ERIS J-band and the optical iron abundances obtained in \citet[red diamonds]{kovtyukh22}, \citet[light blue points]{trentin24A}, and \citet[blue points]{trentin26}. The gray region highlight the uncertainty on the linear fit, displayed with the red line. The functional form of the linear fit is reported in the lower left part of the panel. The gray dotted line represents the 1-1 relation.}}
    \label{fig:comp_literature}
\end{figure}

To ensure the robustness of this method at low-resolution, we collected all of the information from optical high-resolution spectroscopy available in the literature for our targets. First, we compared our metallicities with three literature samples: \citet[8 common CCs]{kovtyukh22}, \citet[10 common CCs]{trentin24A}, and \citet[5 common CCs]{trentin26}. Figure \ref{fig:comp_literature} displays this comparison and a slight negative trend. We performed a linear fit to quantify this trend, obtaining a slope of $0.66\pm0.17$ and an intercept of $-0.01\pm0.03$. The slope is consistent with the 1--1 relation at 2.0~$\sigma$ level. The weighted mean of the residuals $\mathrm{\Delta[Fe/H]}$ (ERIS--literature) is $-0.037\pm0.064$. Considering the typical uncertainties of $\approx$0.1--0.2 dex, we rule out the possibility of strong trends or offsets. Finally, the majority of the comparison is consistent within uncertainties

\begin{figure}
    \centering
    \includegraphics[width=0.485\linewidth]{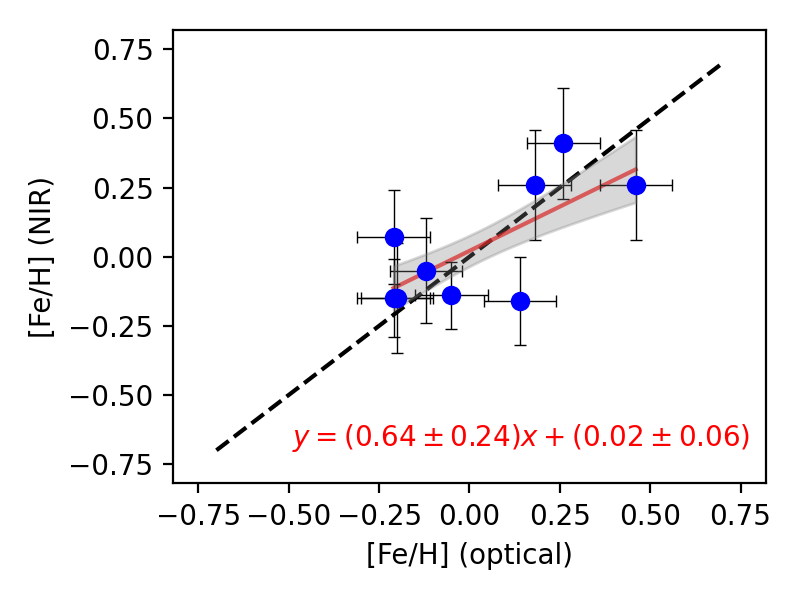}
    \includegraphics[width=0.485\linewidth]{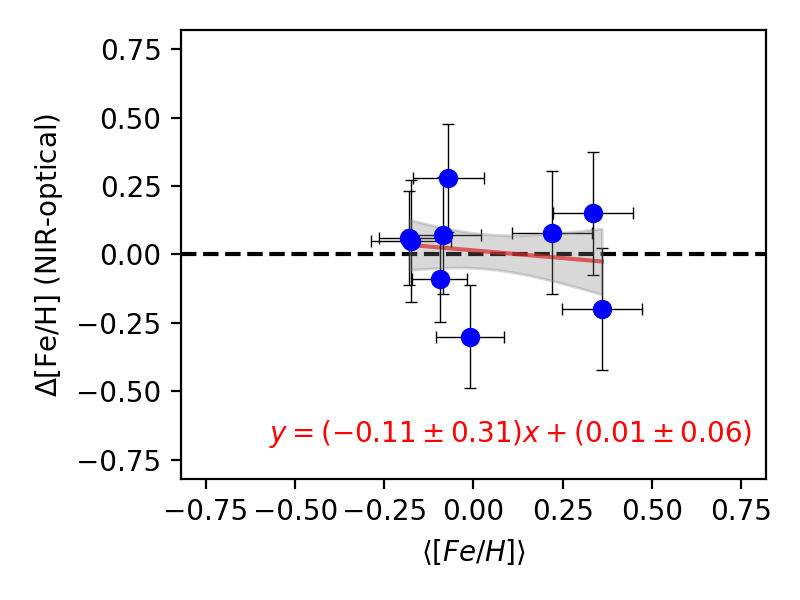}
\includegraphics[width=0.485\linewidth]{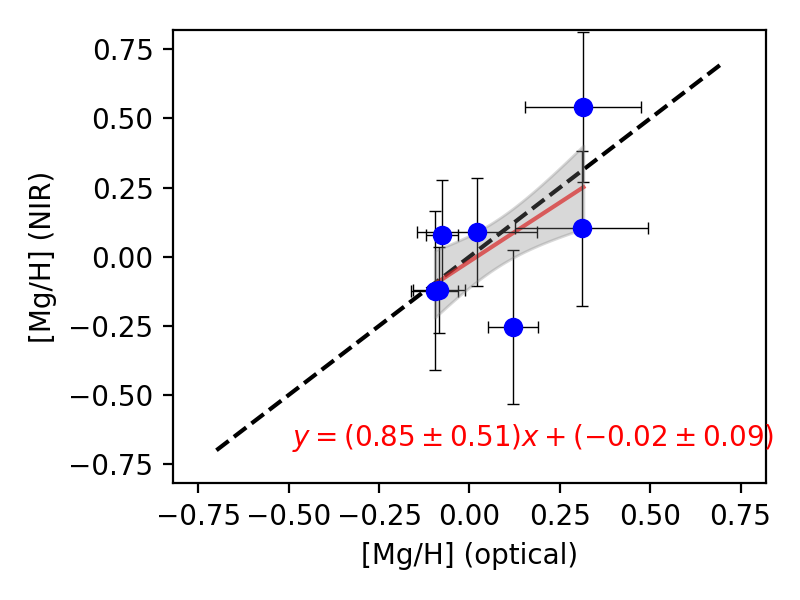}
\includegraphics[width=0.485\linewidth]{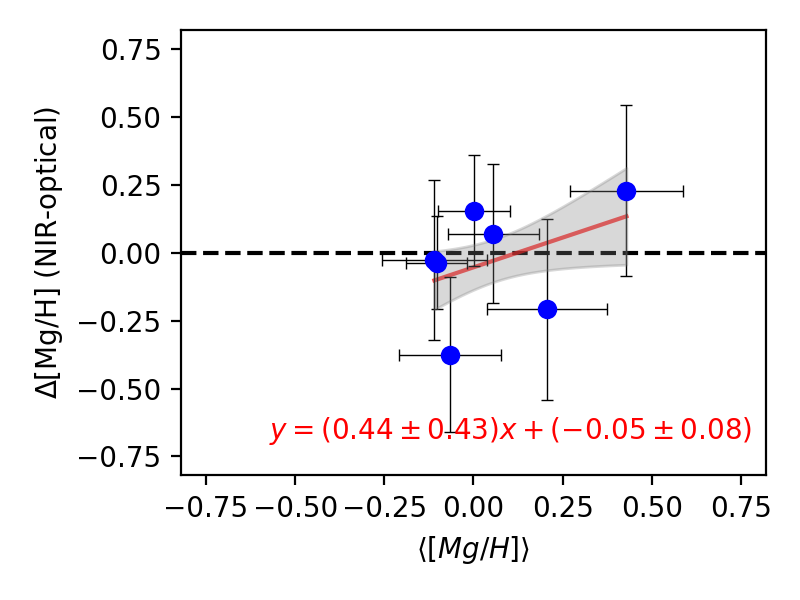}
    \includegraphics[width=0.485\linewidth]{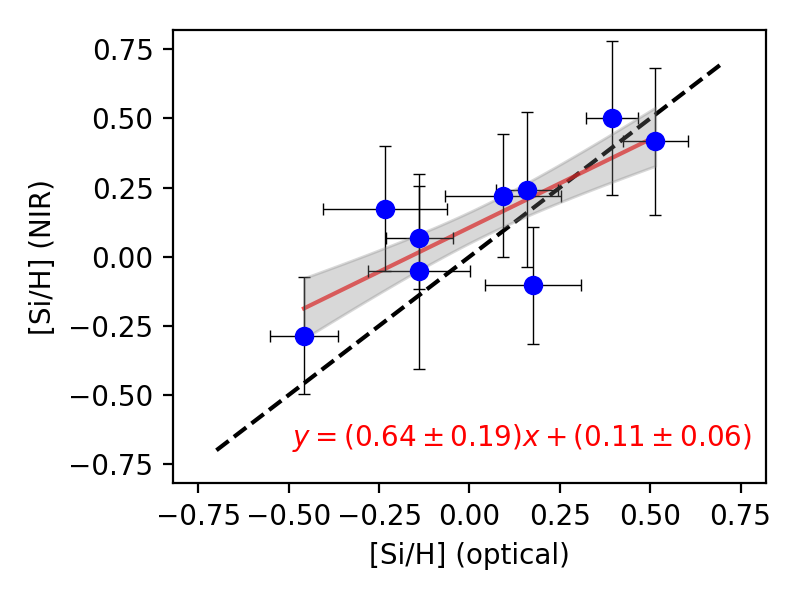}
    \includegraphics[width=0.485\linewidth]{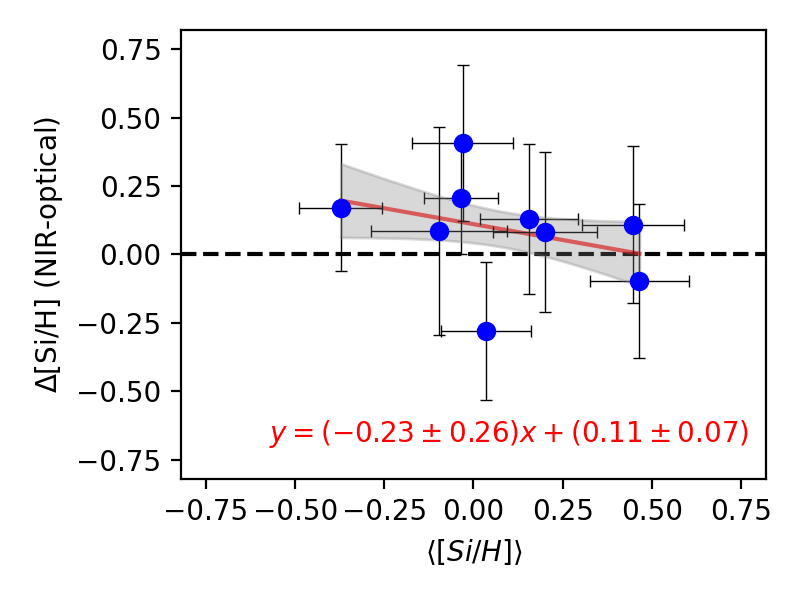}

    \caption{\small{Left panels: Comparisons between NIR J-band and optical abundances. Top panel shows the NIR $\mathrm{[Fe/H]}$ abundances as a function of their optical counterparts, while the middle and bottom panels show the corresponding comparisons for $\mathrm{[Mg/H]}$ and $\mathrm{[Si/H]}$, respectively. The red solid line represents the best-fitting linear relation, and the gray shaded region indicates its uncertainty. The functional form of the linear fit is reported at the bottom of each panel. The gray dotted line represents the 1-1 relation. Right panels: Residuals as a function of the mean abundance, defined as the arithmetic mean of the NIR and optical estimates, for Fe, Mg and Si from top to bottom, respectively. The gray dotted line represents null residuals.}}
    \label{fig:comp_opt_nir}
\end{figure}

Furthermore, we collected 13 optical high-resolution spectra (12 UVES and 1 HARPS) available in the ESO database\footnote{\url{https://archive.eso.org/eso/eso\_archive\_main.html}} for nine of our targets, and estimated the atmospheric parameters following the spectral synthesis method presented in \citet{nunnari26}. Figure \ref{fig:comp_opt_nir} compares the NIR and optical abundances of Fe, Mg and Si (left panels) and shows the differences $\mathrm{\Delta[X/H]~(NIR-optical)}$ as a function of the mean of the two measurements (right panels), which allow us to test for systematics without the correlation introduced by plotting the difference against one of the two quantities.

For [Fe/H], a linear fit of the NIR versus optical values yields a slope of $0.64\pm0.24$ and an intercept of $0.02\pm0.06$. The slope is consistent with unity at the 1.5~$\sigma$ level and the intercept with zero at 0.3~$\sigma$. The $\mathrm{\Delta[Fe/H]}$ versus mean $\mathrm{[Fe/H]}$ relation shows no significant dependence on metallicity: the slope ($-0.11\pm0.31$) is consistent with zero at 0.4~$\sigma$, and the intercept ($0.01\pm0.06$) at 0.2~$\sigma$. For $\mathrm{[Mg/H]}$, the NIR-optical fit has a slope of $0.85\pm0.51$ and an intercept of $-0.02\pm0.09$, both consistent with a one-to-one relation and no offset. The residuals display a slightly positive slope ($0.44\pm0.43$, 1.0~$\sigma$) and an intercept of $-0.05\pm0.08$ (0.6~$\sigma$). Both are consistent with zero, but the large uncertainties on the slope (comparable to the slope itself) mean that the data cannot constrain any trend. For $\mathrm{[Si/H]}$, the NIR versus optical fit gives a slope of $0.64\pm0.19$ (1.9~$\sigma$ from unity) and an intercept of $0.11\pm0.06$ (1.8~$\sigma$ from zero). The residuals show a slope of $-0.23\pm0.26$ (0.9~$\sigma$) and an intercept of $0.11\pm0.07$ (1.6~$\sigma$). Although the NIR $\mathrm{[Si/H]}$ values lie systematically above the optical ones in the metal-poor regime, neither the slope nor the offset departs significantly from zero.

In all three elements, the trends in the right panels are not statistically significant: the uncertainties on the fitted slopes are comparable to or larger than the slopes themselves. For Si, the weighted mean of $\mathrm{\Delta[Si/H]}$ is $+0.094\pm0.089$~dex, consistent with zero at 1.1~$\sigma$. However, we stress that random precision and systematic accuracy are distinct issues. The scatter of the differences is consistent with the random uncertainties of the NIR measurements ($\approx0.2$~dex), but this does not constrain a possible zero-point difference. With N = 9 and the present uncertainties, our test is sensitive only to offsets of a few tenths of dex, and a Si offset of  $\approx0.1$ dex cannot be excluded. Finally, we also report the weighted means for the residuals in Fe and Mg, which are, respectively, $0.001\pm0.066$ and $-0.008\pm0.093$.

Our determined optical and NIR parameters for these CCs are listed in Table~\ref{tab:comp_opt_nir}. Further metallicity comparison with other literature studies that employ data-driven or photometric methods is presented in Appendix~\ref{12ccs}. Finally, further discussion about the atmospheric parameters and related comparison with literature are presented in Appendix~\ref{App:atm}.

\begin{sidewaystable}[ht]
\caption{\small{Optical and NIR atmospheric parameters estimated in the current study.}}
\centering
\resizebox{\textheight}{!}{
    \begin{tabular}{lrrrrrrrrrrrrrrrrrrrrrll}
    \hline
    \hline
Gaia DR3 ID & $\mathrm{T_{eff}}$ & $\mathrm{\log g}$ & $\mathrm{[M/H]}$ & e$\mathrm{[M/H]}$ & $\mathrm{v_{mic}}$ & $\mathrm{[Mg/Fe]}$ & e$\mathrm{[Mg/Fe]}$ & $\mathrm{[Si/Fe]}$ & e$\mathrm{[Si/Fe]}$ & $\mathrm{T_{eff}}$ & e$\mathrm{T_{eff}}$ & $\mathrm{\log g}$ & e$\mathrm{\log g}$ & $\mathrm{[M/H]}$ & e$\mathrm{[M/H]}$ & $\mathrm{v_{mic}}$ & e$\mathrm{v_{mic}}$ & $\mathrm{[Mg/Fe]}$ & e$\mathrm{[Mg/Fe]}$ & $\mathrm{[Si/Fe]}$ & e$\mathrm{[Si/Fe]}$ & Spectrum\_ID & Instr. \\
 & $\mathrm{[K]}$ & $\mathrm{[dex]}$ & $\mathrm{[dex]}$ & $\mathrm{[dex]}$ & $\mathrm{[km\,s^{-1}]}$ & $\mathrm{[dex]}$ & $\mathrm{[dex]}$ & $\mathrm{[dex]}$ & $\mathrm{[dex]}$ & $\mathrm{[K]}$ & $\mathrm{[K]}$ & $\mathrm{[dex]}$ & $\mathrm{[dex]}$ & $\mathrm{[dex]}$ & $\mathrm{[dex]}$ & $\mathrm{[km\,s^{-1}]}$ & $\mathrm{[km\,s^{-1}]}$ & $\mathrm{[dex]}$ & $\mathrm{[dex]}$ & $\mathrm{[dex]}$ & $\mathrm{[dex]}$ & & \\
 & opt & opt & opt & opt & opt & opt & opt & opt & opt & NIR & NIR & NIR & NIR & NIR & NIR & NIR & NIR & NIR & NIR & NIR & NIR & opt & opt \\

\hline
5255092670818771968 & 5890 & 1.62 & $-$0.21 & 0.02 & 4.4 & 0.07 & 0.17 & $-$0.22 & 0.10 & 4750 & 199 & 0.29 & 0.74 & -0.15 & 0.14 & 3.0 & 0.3 &  &  & $-$0.14 & 0.16 & ADP.2024-03-12T09\_19\_50.467.fits & UVES \\
5258418929684479488 & 5729 & 1.71 & $-$0.05 & 0.02 & 3.4 & $-$0.03 & 0.07 & $-$0.06 & 0.10 & 5594 & 198 & 1.91 & 0.80 & $-$0.14 & 0.12 & 2.2 & 0.3 & 0.02 & 0.10 & 0.21 & 0.14 & ADP.2020-10-27T14\_38\_46.471.fits & UVES \\
5336957839849307520 & 5040 & 0.99 & $-$0.12 & 0.02 & 3.7 & 0.04 & 0.04 & $-$0.04 & 0.17 & 6250 & 87 & 2.08 & 0.80 & $-$0.05 & 0.19 & 3.4 & 0.6 & 0.13 & 0.06 & 0.22 & 0.12 & ADP.2020-06-09T15\_05\_32.664.fits & UVES \\
5626365410677889792 & 5656 & 1.01 & $-$0.21 & 0.02 & 4.3 & 0.23 & 0.17 & 0.3 & 0.16 & 6844 & 118 & 2.51 & 0.80 & 0.07 & 0.17 & 4.9 & 0.6 & 0.02 & 0.09 & 0.15 & 0.14 & ADP.2022-05-20T14\_54\_59.318.fits & UVES  \\
 & & & & & & & & & & & & & & & & & & & & & & ADP.2022-05-20T14\_54\_59.329.fits &  \\
5853901851624637824 & 5850 & 1.71 & 0.26 & 0.02 & 3.5 & 0.06 & 0.16 & 0.13 & 0.07 & 6809 & 55 & 2.66 & 0.8 & 0.41 & 0.20 & 3.5 & 0.6 & 0.13 & 0.10 & 0.09 & 0.12 & ADP.2020-10-27T14\_38\_46.504.fits & UVES \\
5862711040048310528 & 5232 & 0.88 & 0.14 & 0.02 & 3.5 & $-$0.02 & 0.07 & 0.04 & 0.13 & 4865 & 112 & 2.00 & 0.80 & $-$0.16 & 0.16 & 3.3 & 0.3 & $-$0.09 & 0.23 & 0.06 & 0.14 & ADP.2024-03-27T09\_10\_11.648.fits & UVES \\
5862720214099407232 & 5194 & 0.53 & $-$0.20 & 0.02 & 5.6 & 0.10 & 0.06 & 0.06 & 0.14 & 4750 & 350 & 0.80 & 0.43 & $-$0.15 & 0.20 & 3.6 & 0.4 & 0.03 & 0.14 & 0.10 & 0.26 & ADP.2024-03-27T09\_04\_07.352.fits & UVES \\
5880087446907041152 & 5156 & 0.99 & 0.18 & 0.02 & 3.8 & $-$0.04 & 0.09 & $-$0.02 & 0.08 & 4985 & 345 & 1.69 & 0.80 & 0.26 & 0.20 & 2.4 & 0.4 &  &  & $-$0.02 & 0.12 & ADP.2020-10-27T14\_38\_46.490.fits & UVES \\
5893788220810031872 & 6201 & 1.31 & 0.46 & 0.08 & 4.6 & $-$0.15 & 0.18 & 0.05 & 0.08 & 5517 & 350 & 1.17 & 0.26 & 0.20 & 0.25 & 4.4 & 0.6 & $-$0.16 & 0.12 & 0.16 & 0.09 & ADP.2024-03-27T09\_04\_07.344.fits & UVES \\
\hline
\hline
\end{tabular}
 }
\tablefoot{Information for each optical-NIR comparison, including the atmospheric parameters and chemical abundances. The Gaia DR3 source ID is listed in the first column. Columns (2)-(10) contain the optical atmospheric parameters and chemical abundances, together with abundance uncertainties. Columns (11)-(22) contain the NIR atmospheric parameters and chemical abundances, together with their respective uncertainties. Columns (23) and (24) show the optical spectrum ID and the corresponding instrument, respectively.} 
\label{tab:comp_opt_nir}
\end{sidewaystable}

\begin{figure}
    \centering
    \includegraphics[width=0.32\linewidth]{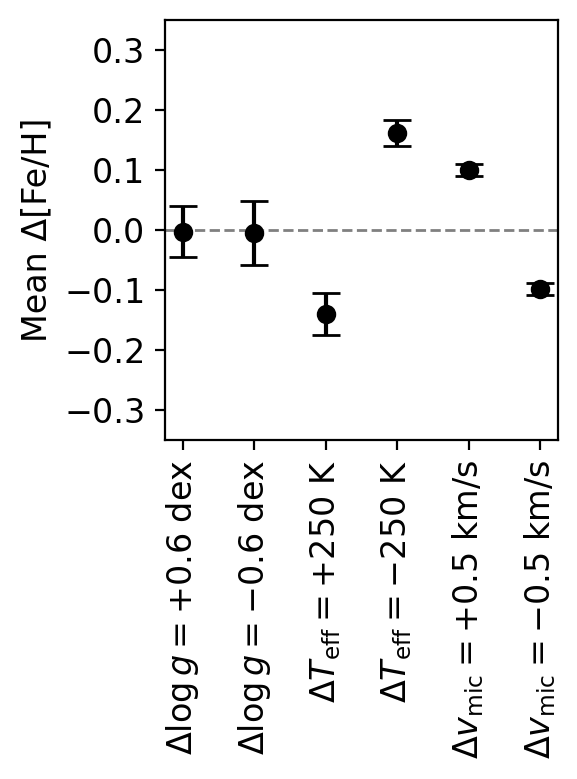}
    \includegraphics[width=0.32\linewidth]{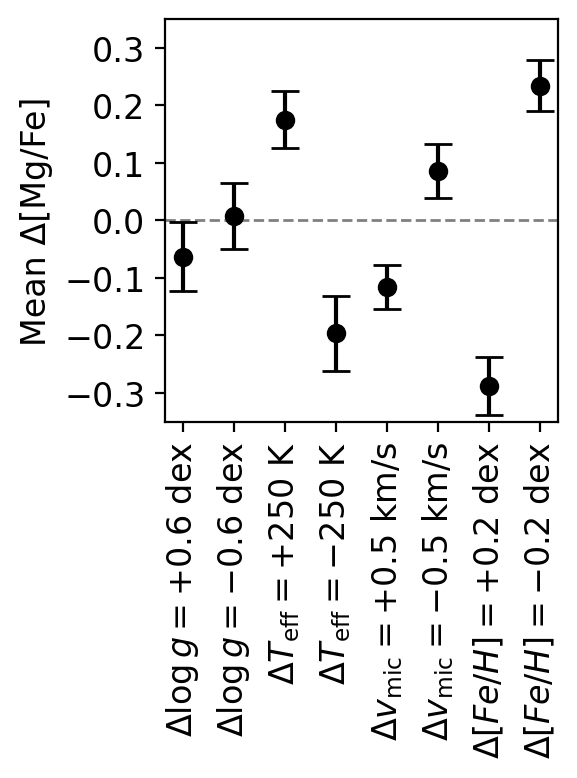}
    \includegraphics[width=0.32\linewidth]{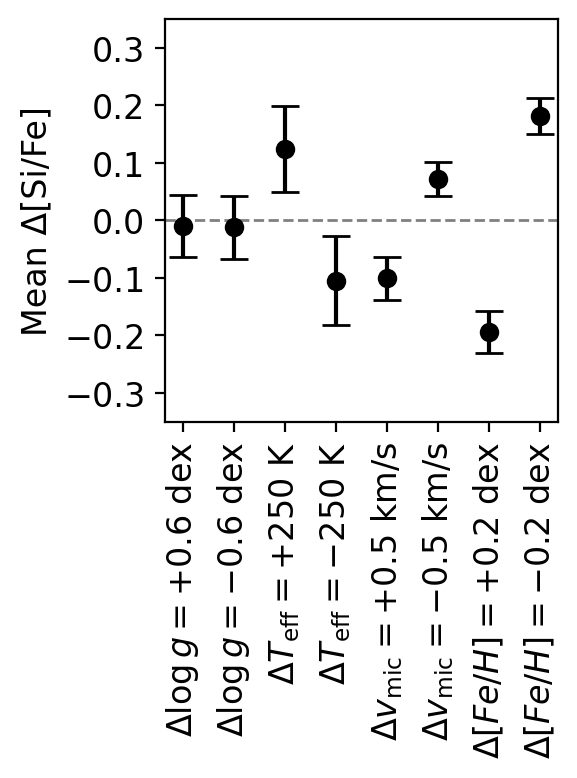}

    \caption{\small{Sensitivity of the iron abundance (left panel), Mg ratios (central panel) and Si ratios (right panel) to variations in $\mathrm{\log g}$, $\mathrm{T_{eff}}$ and $\mathrm{v_{mic}}$. For Mg and Si we also show the sensitivity to variations of $\mathrm{[Fe/H]}$. The y-axis represents the mean difference between the $\mathrm{[Fe/H]}$ or abundance ratio derived with varied parameters and the value obtained with the original parameters. The error bars are the standard deviations with respect to the mean differences.}}
    \label{fig:sens_params}
\end{figure}

We find average uncertainties of 200--300~K in $\mathrm{T_{eff}}$, 0.5--0.7~dex in $\mathrm{\log g}$ and 0.4--0.6~$\mathrm{km \,\, s^{-1}}$ in $\mathrm{v_{mic}}$. To quantify the impact of these uncertainties on the derived abundances, we performed a sensitivity analysis by varying $\mathrm{T_{eff}}$ $(\pm250~\mathrm{K})$, $\mathrm{\log g}$ ($\pm0.6$~dex) and $\mathrm{v_{mic}}$ $(\pm0.5~\mathrm{km~s^{-1}})$. The results, shown in Fig.~\ref{fig:sens_params}, indicate that effective temperature is the most influential parameter affecting the metallicity and, on the contrary, surface gravity is the least influential. 
In the central and right panel of Fig.~\ref{fig:sens_params} we show the sensitivity of $\mathrm{[Mg/Fe]}$ and $\mathrm{[Si/Fe]}$ to variations of the same parameters as previously described and in global metallicity ($\pm$0.2~dex), finding that these chemical ratios exhibit consistent responses that do not compromise the general chemical trends. 

\subsection{Distance estimation}\label{sec:distance}

To inspect the gradients of chemical abundances and locate our sample of CCs within the MW, we estimated their individual distances, based on different diagnostics: 1) W1-band Period-Luminosity relation (PL, calibrated from Galactic CCs by \citealt{wang18}); 
2) Ks-band PL \citep[calibrated on Galactic CCs by][]{ripepi20}; 
3) G/Bp/Rp Period-Wesenheit relation (PW, calibrated from Galactic CCs by \citealt{ripepi19}).

We did not adopt any distance based on Gaia parallaxes in order to avoid introducing biases in our estimates. In fact, our quality cuts for Gaia astrometry leave us with only 15 Cepheids and, most importantly, all within 5~kpc. PLs and PWs, on the other hand, allow distance measurements on a much broader dynamical range, including all our Cepheids. In addition, we did not adopt metallicity-dependent PLs and PWs, although \citet{trentin24A} published an extensive set of relations, because there is no solid consensus on the value of the metallicity coefficient in the PL relations, with values ranging as much as 0.5~mag/dex \citep{breuval21, trentin23, bono2024cepheids}. Moreover, there are no calibrations of metallicity-dependent PLs in the WISE passbands, so we would introduce a bias between MIR-based and NIR-based distance estimates. Nevertheless, the impact of metallicity in NIR PLs is expected to be weaker (\citealt{fiorentino07, fiorentino13, ngeow12, dicriscienzo13, gieren18, ripepi20, trentin23}). 

Mid- and Near-Infrared magnitudes (MIR, W1/W2 and NIR, J/Ks) were adopted from the WISE and 2MASS surveys, respectively, adopting the cross-match performed by \citet{marrese19} and publicly available on the SSDC Gaia portal\footnote{\url{https://gaiaportal.ssdc.asi.it/DR3/home}}. However, we only adopted magnitudes for which the photometric quality (\texttt{PhotQual} for WISE and \texttt{PSCphQual} for 2MASS) is either A, B or C, discarding bad quality measurements. Concerning the Gaia magnitudes (G, Bp, Rp) we only selected those with uncertainty smaller than 0.2~mag, meaning less than 20\%  uncertainty in flux.

For the distance estimates from PLs, the MIR and NIR magnitudes must be unreddened, adopting a reddening map and a reddening law. Concerning the reddening map, we adopted as a ground-level solution, the reddening map by \citet{schlegel98}, rescaled by \citet{schlafly11}. However, when $\mathrm{E(B-V)}$ exceeds 2 mag, we instead adopted the extinction in the G band ($A_G$) derived within Gaia DR3 from the Gaia GSP pipeline (\texttt{ag\_gspphot} column, \citealt{andrae23}). Also these values may be flawed, therefore we converted the $A_G$ values into $\mathrm{E(B-V)}$ adopting the reddening law \citet{wang19}. If the equivalent $\mathrm{E(B-V)}$ was larger than 2~mag, we discarded also the Gaia-based values. However, our sample also includes stars towards the Galactic Bulge, which is a peculiar region concerning reddening, with both higher values of $\mathrm{E(B-V)}$ and a more pronounced variation of $\mathrm{E(B-V)}$ on smaller angular scales. For this reason, we adopted the $\mathrm{E(J-K)}$ values by \citet{surot20}, based on NIR VVV photometry, for CCs within area covered by the map. Concerning the reddening law, we adopted the \citet{cardelli89} law, modified as suggested by \citet{stetson14}, for the J/Ks passbands and the \cite{madore13} law for the MIR magnitudes. For the J/K magnitudes of CCs within the \citet{surot20} map, we adopted the reddening law by \citet{alonsogarcia17}, which is specific for the inner regions of the MW.

PWs, where the Wesenheit magnitude is G~$-$~1.9~(Bp~$-$~Rp) are, instead, reddening-independent by construction, provided that the reddening law adopted to estimate the color coefficient (1.9 in our case) is the same for all the target stars. This is of course an unrealistic assumption, but the inadequacy of PWs is significant only in extremely reddened regions like the innermost Bulge. Moreover, since the calibration of the PW is based on Cepheids all over the galactic disk, the intrinsic dispersion of the relation, which is considered in the uncertainty estimate, already includes the effect of a non-uniform reddening law.

Finally, uncertainties are obtained by propagating both the photometric uncertainties, the uncertainties on the coefficients and the intrinsic spread of the relation. We did not propagate the uncertainty on the period because these are tiny ($10^{-6}$--$10^{-5}$, fractionally) and would make an insignificant part of the overall uncertainty.

The adopted hierarchy is: PL(W1) when available; otherwise PL(Ks); if neither W1 nor Ks photometry is available, or if the photometric quality parameter in WISE and 2MASS is lower than "C", or if the reddening is not reliable (when E(B-V)> 2 mag, this is usually inaccurate), we adopt the PW(G, BP–RP) relation. According to this hierarchy, we ended up with distances for, respectively, 15, 2 and 10 stars with individual uncertainties ranging from 3 to 13\%.

\section{Results and Discussion}
\label{sec:results}
\begin{figure}
    \centering
    \includegraphics[width=1.\linewidth]{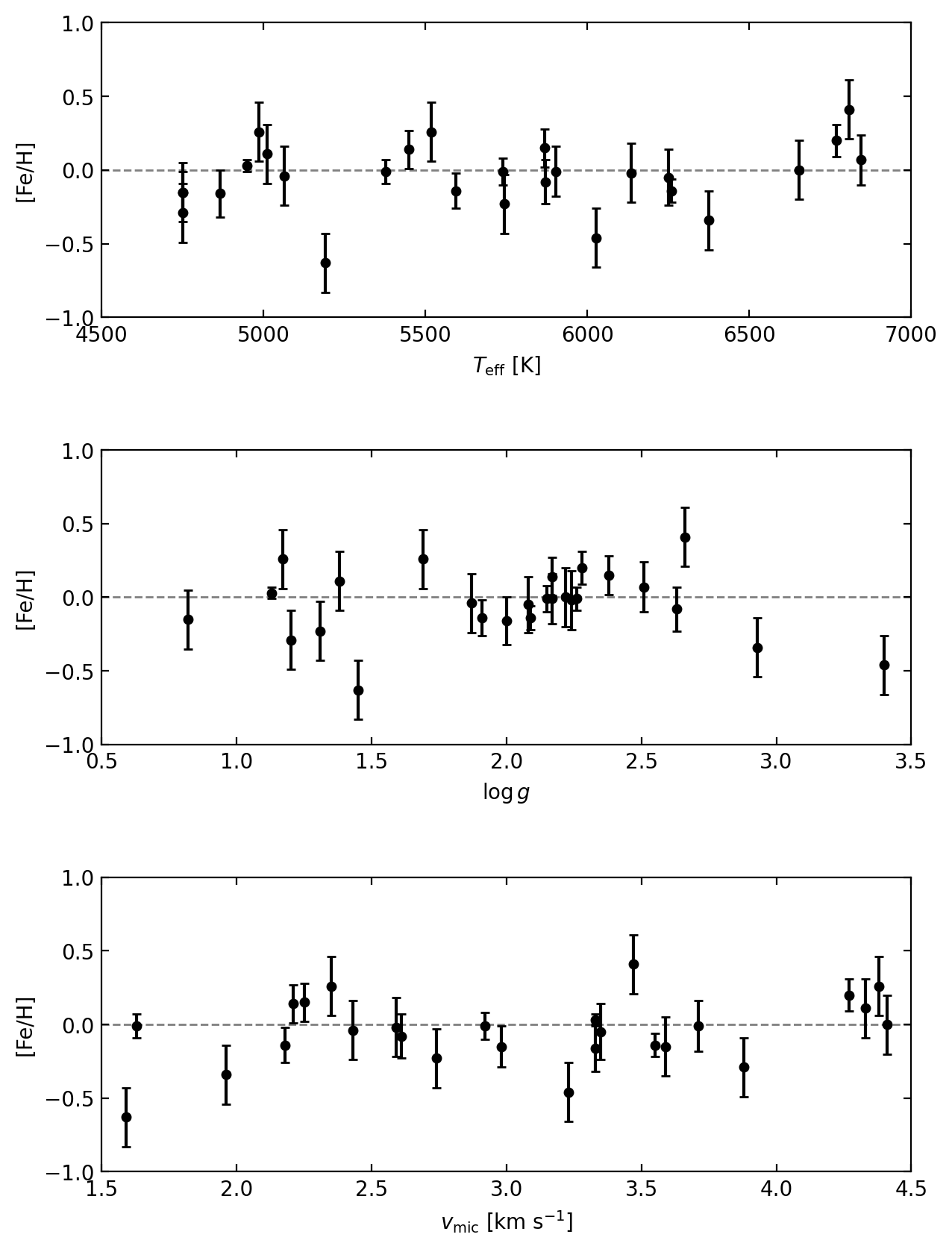}
    \caption{\small{Iron abundances as a function of $\mathrm{T_{eff}}$ (top panel), of $\mathrm{\log g}$ (central panel), and of $\mathrm{v_{mic}}$ (bottom panel) as obtained from our spectral analysis.}}
    \label{fig:trends}
\end{figure}

The atmospheric parameters, chemical abundances and distances of the 27 CCs are listed in Table 3. Our spectral analysis yields a broad range of fundamental parameters, with $\mathrm{T_{eff}}$ spanning from approximately 4800~K to 6800~K and $\mathrm{\log g}$ between 0.5~dex and 3.0~dex, typical ranges of CCs (\citealt{da2023oxygen}; \citealt{bono2024cepheids}; \citealt{markova25}; \citealt{nunnari26}; \citealt{trentin26}). The derived $\mathrm{[Fe/H]}$ from approximately $-$~0.6 to 0.5~dex is within the expected range for Galactic Cepheids. Notably, Figure~\ref{fig:trends} shows that there are no significant trends between metallicity and other atmospheric parameters, supporting the internal consistency of our optimization method. 

Moreover, we provide the first Mg and Si abundances measurements for 12 CCs, while their metallicity was obtained in different previous studies using data-driven applied to Gaia spectra or photometric approach. We discuss in detail and show the metallicity comparison between these studies and the current work in Appendix~\ref{12ccs}. CCs with the first Mg and Si abundances measurements are flagged with "new\_alpha" in the column "new target" of Table 3.  
Particularly, the object with Gaia DR3 ID 5415098722470316544 is located at a Galactocentric distance of 28.1~$\pm$~0.3 kpc, in a region of the disk still under-sampled, with a notable height of $\mathrm{Z}$~=~+4.8~$\pm$~1.1kpc. 

\subsection{Kinematic properties}
\begin{figure*}
    \centering
    \includegraphics[width=1.\linewidth]{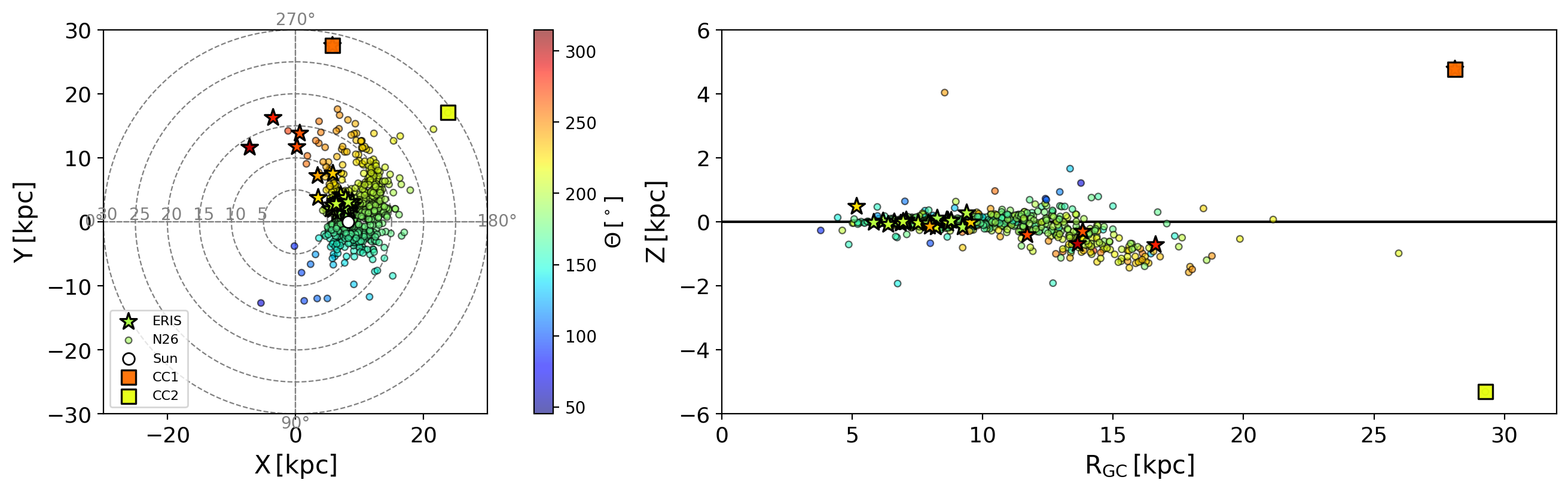}
    \caption{\small{Distribution of the CCs across the thin disk (left panel) and the height above the Galactic plane (Z) as a function of $\mathrm{R_{GC}}$ (right panel), both color-coded according to the azimutal angle. The star  symbols are used for this study sample, while the points are employed for the \citet{nunnari26} kinematic sample. CC1 and CC2 are displayed respectively with the red and the orange squares.}}
    \label{fig:Z_vs_RGC}
\end{figure*}

Figure~\ref{fig:Z_vs_RGC} shows the distribution of the CCs across the thin disk color-coded according to the azimutal angle (left panel) and the height above the Galactic plane (Z) as a function of $\mathrm{R_{GC}}$ (right panel). The ERIS sample follows the trend in Z vs $\mathrm{R_{GC}}$ similarly to that of \citet{nunnari26}, where the decreasing trend toward negative values of Z, which occurs at around 12~kpc, is an effect of the warp of the Galactic disk (see also Fig.~A.3 of \citealt{lemasle22}; \citealt{minniti21}). However, the Cepheid "Gaia DR3 5415098722470316544" (hereafter CC1) has $\Theta\simeq260^{\circ}$ and an extreme value of Z~=~+4.8~$\pm$~1.1~kpc. At this azimut, the warped disk shows on average negative values of Z (\citealt{zhou25, poggio26}). In addition, 4 targets of the ERIS sample have $\mathrm{R_{GC}}$~>~10~kpc, in a region where the disk flare becomes not negligible. The flare of the outer disk, which consists of an increase in its vertical scale height compared with the inner disk, is supported both from numerical simulations (\citealt{minchev12,minchev14,minchev15}) and observational evidence (\citealt{c22, lemasle22, poggio25, poggio26, zhou25}), mapping up to $\mathrm{R_{GC}\approx}$~20~kpc. 

To establish that our targets belong to the disk, we used the software package \textit{AGAMA} (\citealt{AGAMA}) together with the newest galactic potential from \citet{Hunter2024}, which includes contributions from the supermassive black hole Sgr~$\rm A^{*}$, the nuclear star cluster, the nuclear stellar disk, the Galactic bar, the Galactic disk including the spiral arms and the dark matter halo. With Gaia astrometry, radial velocities, and distances from Sect. \ref{sec:distance}, we integrated the orbits over  period of 13~Gyr with a time step of 10~Myr. In addition, we have implemented Markov Chain Monte Carlo (MCMC) simulations (100) taking into account the distance uncertainties. We adopted a Solar Galactocentric distance of $\mathrm{R_{\odot}}=8.2$~kpc (\citealt{blandhawthorn16review}) and a Solar height above the Galactic plane of $\mathrm{Z_{\odot}}=23$~pc (\citealt{joshi05}). The Solar motion with respect to the local standard of rest was taken to be (U$_{\odot}$, V$_{\odot}$, W$_{\odot}$)=(11.1, 12.24, 7.25)~$\mathrm{km~s^{-1}}$ (\citealt{schonrich10}).

\begin{figure*}
    \centering
    \includegraphics[width=0.98\linewidth]{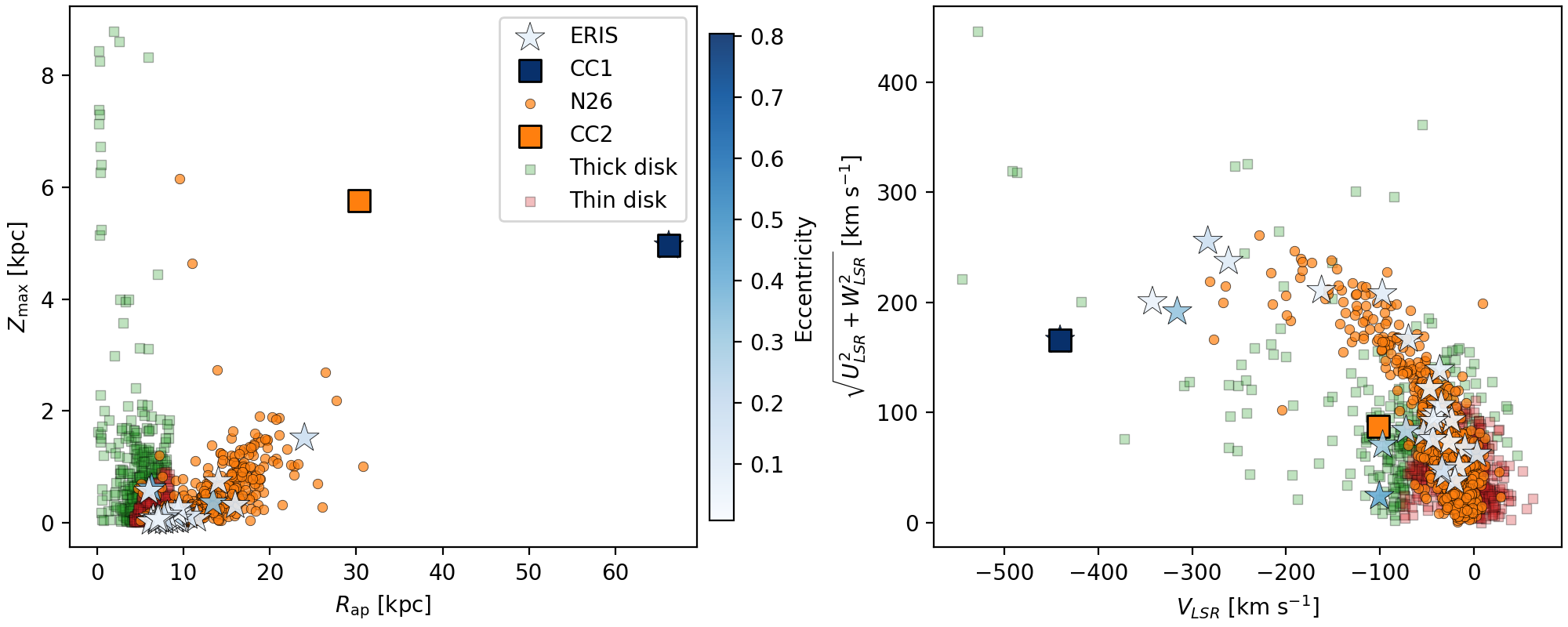}
    \caption{\small{Maximum verical height ($\mathrm{Z_{max}}$) as a function of the apocentric radius ($\mathrm{R_{ap}}$) of the orbit (left panel). The sum in quadrature of $\mathrm{U_{LSR}}$ and $\mathrm{W_{LSR}}$ as a function of $\mathrm{V_{LSR}}$ (right panel, Toomre diagram). $\mathrm{U_{LSR}}$, $\mathrm{V_{LSR}}$, and $\mathrm{W_{LSR}}$ are the components of a star's motion in a galactic Cartesian reference frame centered on the Local Standard of Rest (LSR), which removes the Sun's peculiar motion. The ERIS sample is color-coded by the eccentricty and is depicted with stars symbol, while the \citet{nunnari26} kinematic sample is highlighted with orange circles. Big square symbols are used to distinguish CC1 and CC2. Green and red small squares respectively represent the thick (TD/D>1) and thin (TD/D<1) disk population from \citet{bensby+14} as a reference.}}
    \label{fig:agama}
\end{figure*}

Figure~\ref{fig:agama} shows two important kinematic planes: maximum vertical height ($\mathrm{Z_{max}}$) as a function of the apocentric radius ($\mathrm{R_{ap}}$) in the left panel and the Toomre diagram in the right panel. Most of the sample lies on typical thin disk orbits given their low eccentricity (<0.3) and high circularity ($\mathrm{\lambda_{z}~>~0.8}$). The fraction of the sample at $\mathrm{R_{GC}~>~10~kpc}$, which is characterised by high $\mathrm{Z_{max}}$ values (0.5~<~$\mathrm{Z_{max}}$~<~1.5~kpc) is affected by the warp. This means that despite the comparison with \citet{bensby+14} suggest a thick disk characterization of these Cepheids, they are consistent with being thin-disk tracers affected by the warp as suggested by Fig.~\ref{fig:Z_vs_RGC} where the warped thin disk is evident. Notice that \citet{bensby+14} sample is limited to about $\mathrm{R_{GC}~<~10~kpc}$ where the warp is negligible. Finally, CC1 has a very eccentric orbit (eccentricity~$=0.8\pm0.1$), an extreme $\mathrm{Z_{max}}$ of $4.96\pm1.40$~kpc and a $\mathrm{R_{ap}}$ of $66\pm6$~kpc.

The fact that CC1 shows halo-like orbits in the kinematic planes casts doubt on the idea that it is tracing the disk. CCs are intermediate-mass stars with relatively young ages (see Appendix~\ref{app:light_curve}), thus suggesting that it might be, as a working hypothesis, a runaway disk star (\citealt{carreterocastrillo23}; \citealt{stoop24}). A more detailed analysis of the kinematic properties of the outermost CCs will be discussed in a forthcoming paper (Pepe et al. in preparation). For this reason, we mark differently CC1 from the rest of the sample in the following sections and figures. It is worth mentioning that the CC at $\mathrm{R_{GC}}\approx29$~kpc (ASAS~J062939-1840.5, hereafter CC2) of the optical sample is characterised by a circular orbit, with eccentricity of $\approx0.23$, unlike CC1 which has a value of $\approx0.80$.
CC2 is characterised by $\mathrm{R_{GC}}~\approx$~29~kpc and $\mathrm{Z}~\approx-$5.3~kpc. Despite the extreme value of Z, Fig. \ref{fig:agama} suggests a disk-like orbit, hinting at the possible extension of the warp of the disk toward more negative values of Z than expected from other studies (CCs, \citealt{poggio26}; HI, \citealt{kalberla07}, \citealt{Soding25}; supergiants, \citealt{c22}; red clump stars, \citealt{uppal23}). 
The reader may wonder whether our method (Sect.~\ref{sec:distance}) could overestimate the distance. For CC2, the StarHorse distance (\citealt{anders+22}), based on a Bayesian combination of Gaia astrometry and multi-band photometry, yields a smaller value that would place CC2 at a more plausible Z ($\approx-$2~kpc). However, this estimate relies on a negative parallax, making it intrinsically unreliable; the PL distance is therefore more trustworthy. This is further supported by \citet{skowron19}, who derived a PL-based distance of $23.6\pm1.2$~kpc, in excellent agreement with our value of $23.7\pm0.3$~kpc. CC2 was also studied in \cite{trentin23}, who instead obtained a PL-based distance of $19.8\pm1.2$~kpc, corresponding to Z~$\approx-$4.4~kpc.

Although this discussion is extremely interesting, it goes beyond the scope of this paper; therefore, we will address it in a future work. However, a further discussion is addressed in Appendix~\ref{app:light_curve}, where we show the light curves of CC1 and CC2, that rule out a possible misclassification of these objects as young classical Cepheids.

\subsection{Abundance ratios: Mg and Si}
\begin{figure}
    \centering
    \includegraphics[width=1.\linewidth]{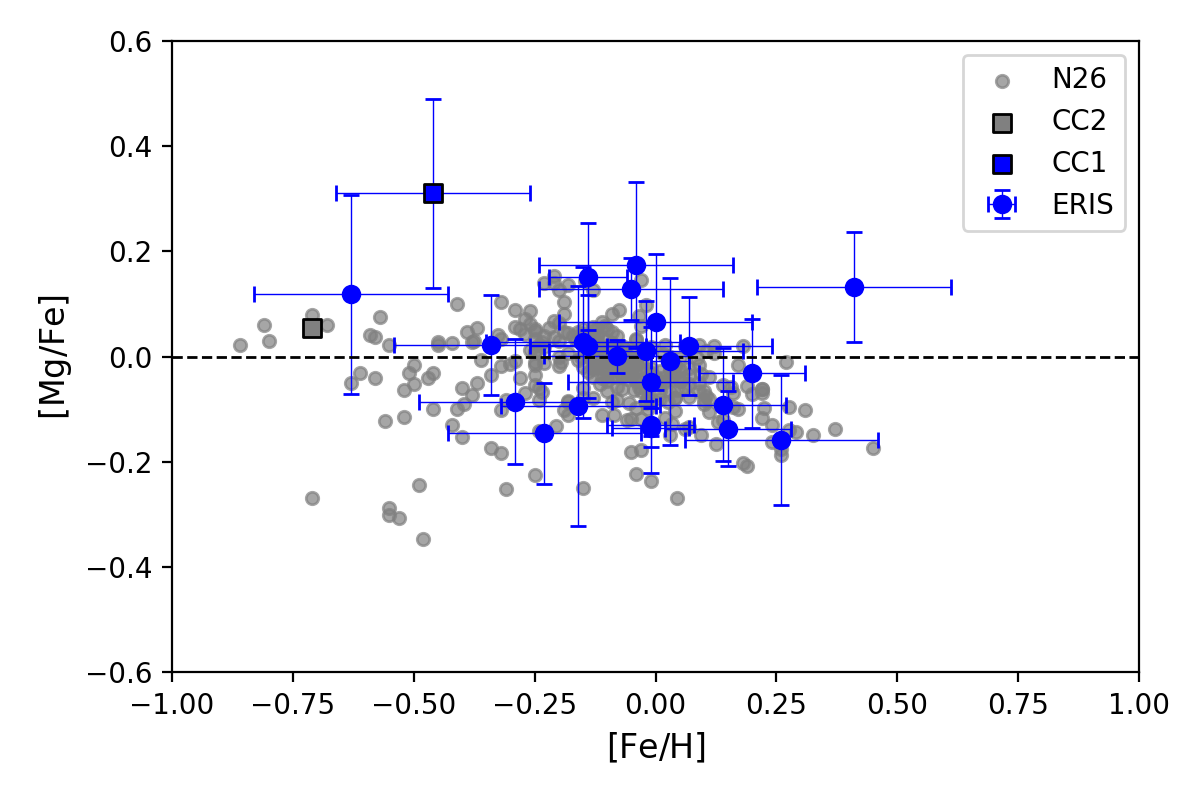}
    \includegraphics[width=1.\linewidth]{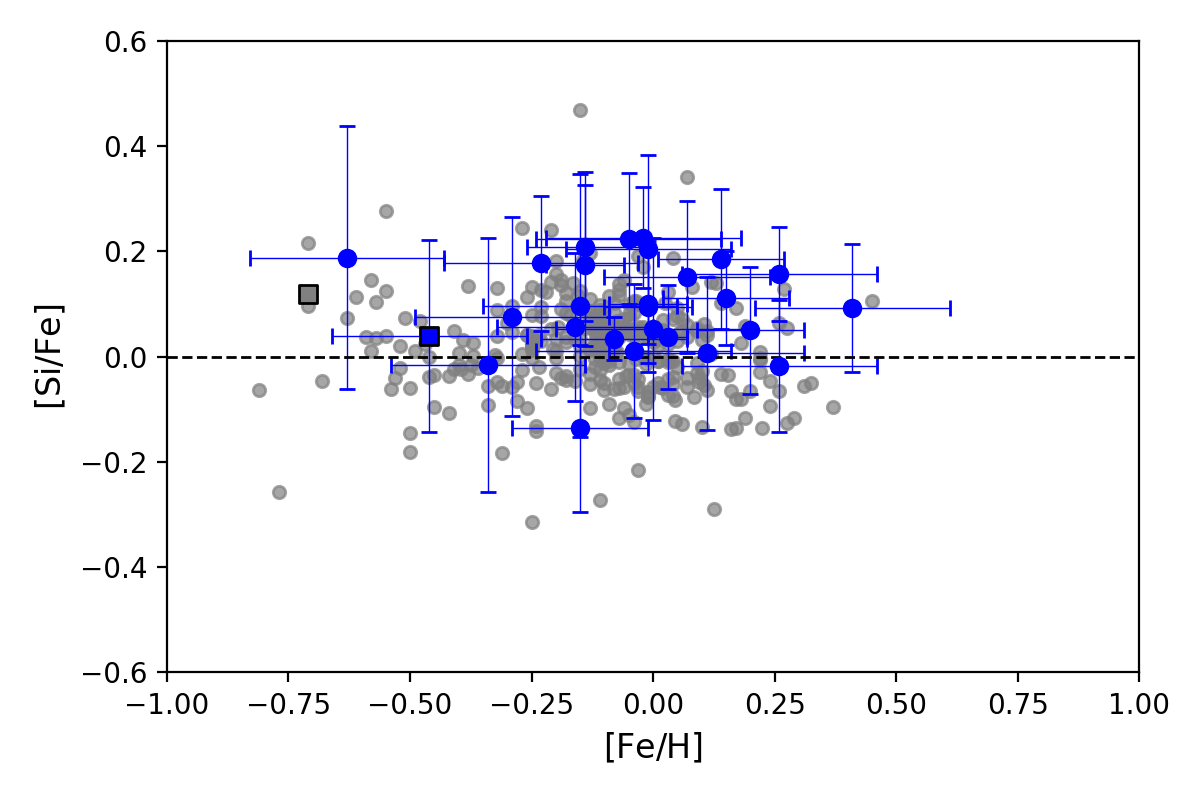}
    \caption{\small{[Mg/Fe] (top panel) and [Si/Fe] (bottom panel) as a function of metallicity. Blue points depict the estimates of the ERIS sample, while gray points those of \citet[N26]{nunnari26}. CC1 and CC2 are shown respectively with the blue and the gray squares.}}
    \label{fig:Mg_Si}
\end{figure}

The abundance ratios of Mg and Si as a function of metallicity are displayed in Fig.~\ref{fig:Mg_Si}. We find mean values and relative standard deviations of $\mathrm{\langle[Mg/Fe]\rangle}=0.01\pm0.12$~dex and $\mathrm{\langle[Si/Fe]\rangle}=0.10\pm0.09$~dex. These results show excellent agreement with the high-resolution optical analysis of 401 Galactic CCs by \citet{nunnari26}, who reported $-0.03\pm0.08$~dex and $0.02\pm0.09$~dex for $\mathrm{[Mg/Fe]}$ and $\mathrm{[Si/Fe]}$, respectively, and with those of the 328 Galactic CCs sample of \citet{trentin26}, which are $+0.14\pm0.19$~dex for Mg and $+0.19\pm0.21$~dex for Si. 

This consistency is significant because magnesium and silicon, which are primarily produced in core-collapse supernovae, show $\mathrm{[X/Fe]}$ ratios relatively flat with increasing iron content across the young Galactic disk (open clusters, \citealt{dalponte25}; CCs, \citealt{trentin26}), aligned with current chemical evolution models that predict nearly solar $\mathrm{[\alpha/Fe]}$ ratios for the young stellar population in the thin disk (\citealt{spitoni23a}). 

The agreement in Mg and Si ratios provides an additional consistency check in favour of the J-band diagnostic power, mirroring findings by \citet{inno19}, \citet{matsunaga23}, \citet{catanzaro26}, who demonstrated that high-resolution NIR spectroscopy for CCs yields robust abundances fully compatible with optical results (e.g., \citealt{catanzaro26}, $\mathrm{\Delta[Fe/H]}~\leq~0.02$~dex). Even at our low resolution (R~$\sim$~4000), the stability of the metallicity and $\alpha$-element abundance estimations remains solid, supporting the conclusion that the J-band is a viable tool for chemical investigation in highly obscured regions (\citealt{davies10}; \citealt{lardo15}).  

\subsection{Chemical gradients}
The spatial distribution illustrated in Fig.~\ref{fig:Z_vs_RGC}, confirms that the ERIS sample is distributed across the Galactic thin disk, spanning $\mathrm{R_{GC}}$ from approximately 5 to 28~kpc, with the bulk concentrated between 5 and 10~kpc. It provides a useful baseline for tracing the chemical properties of the inner, outer, and local disk. 

\begin{figure}
    \centering
    \includegraphics[width=1\linewidth]{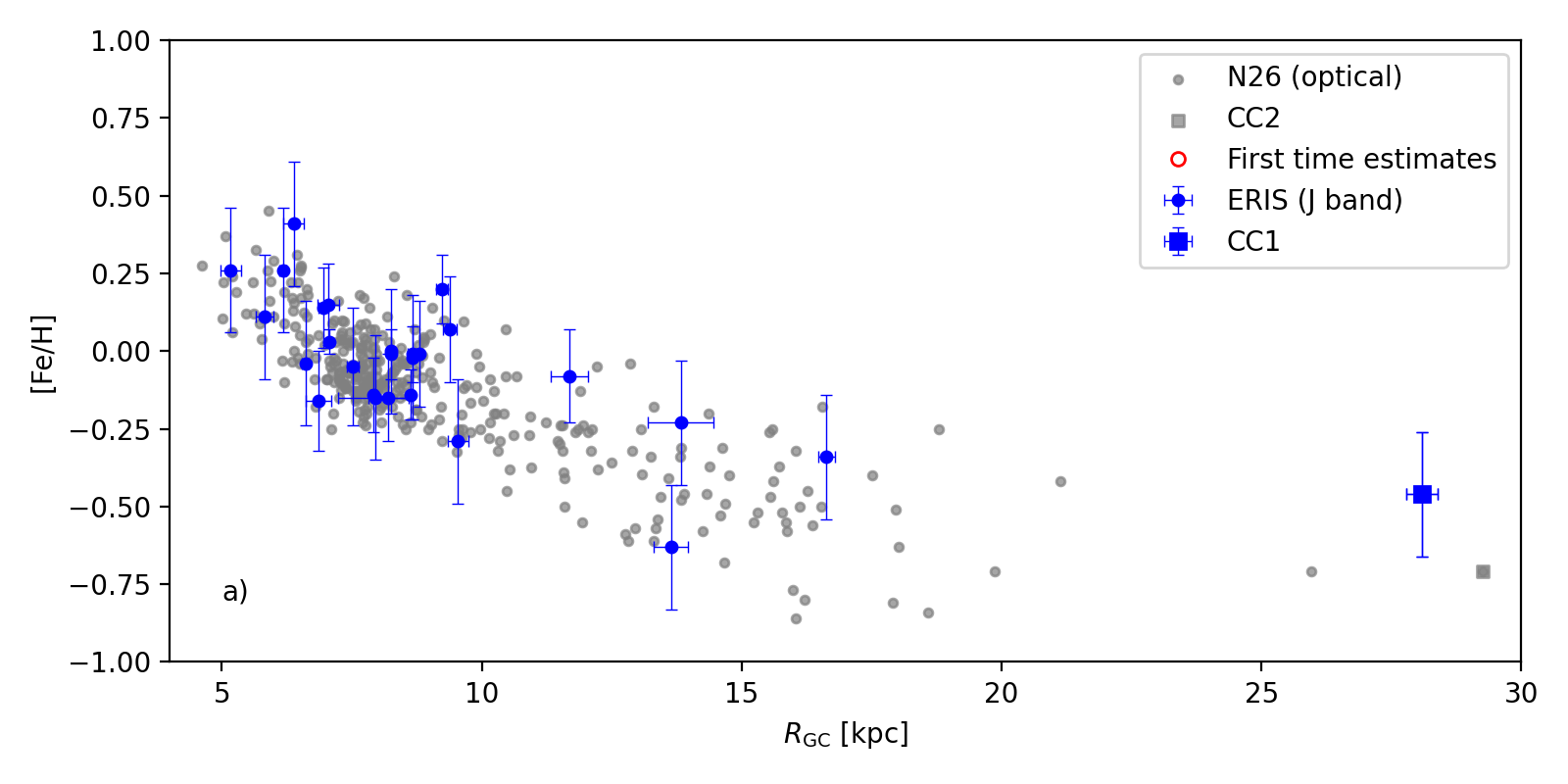}
    \includegraphics[width=1.\linewidth]{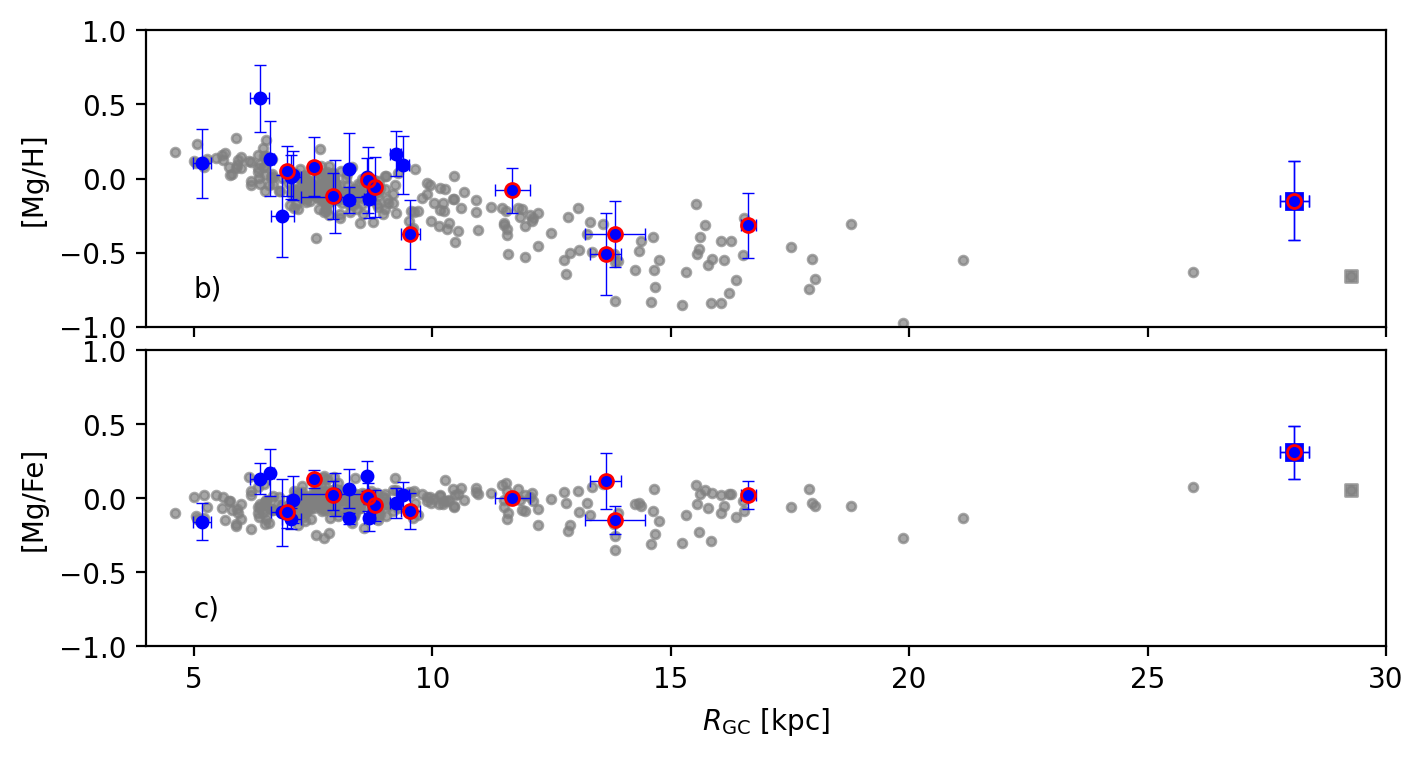}
    \includegraphics[width=1.\linewidth]{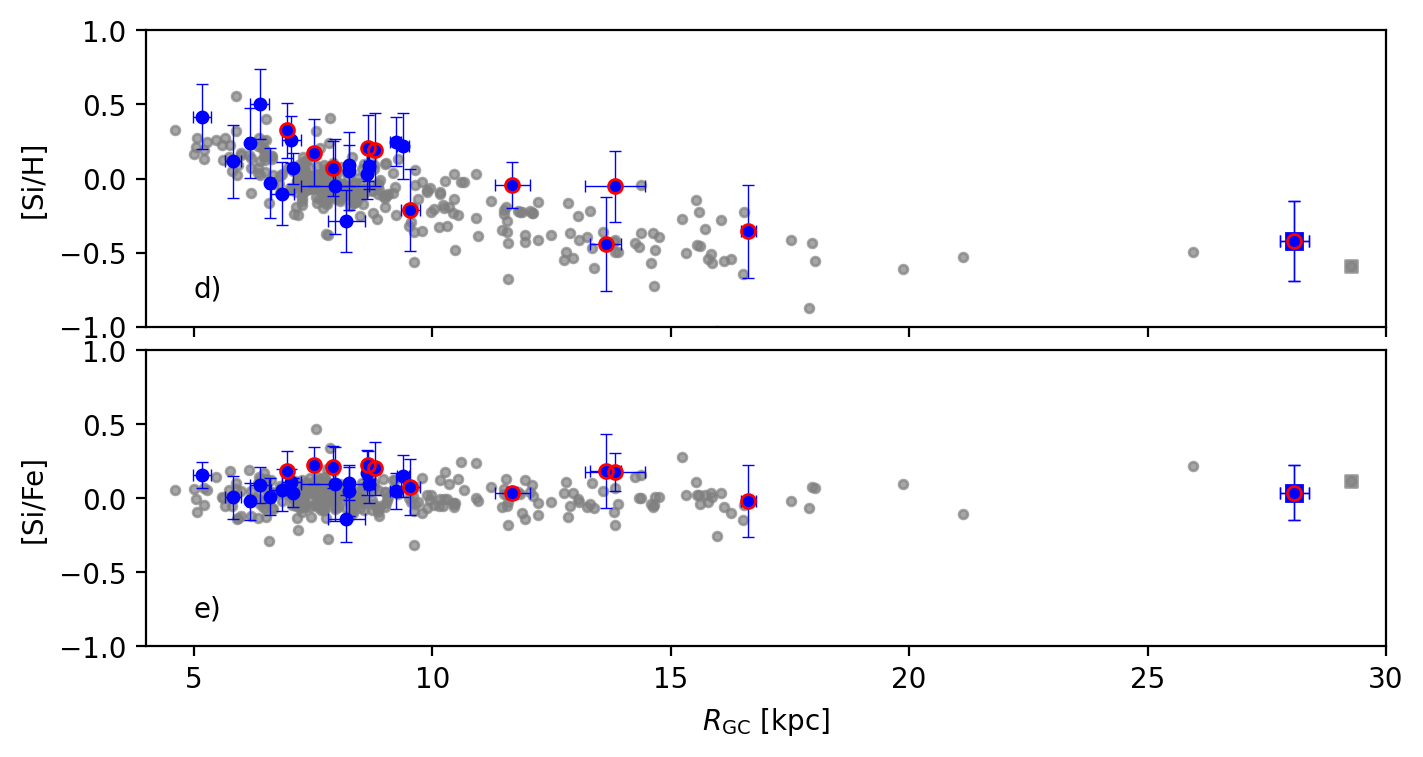}
    \caption{\small{[Fe/H] (panel a), [Mg/H] (panel b), [Mg/Fe] (panel c), [Si/H] (panel d) and [Si/Fe] (panel e) as a function of $\mathrm{R_{GC}}$. The ERIS and the \citet[N26]{nunnari26} samples are shown respectively in blue and in gray. The targets with the first time estimates of Mg and Si abundances are highlighted in red. CC1 and CC2 are displayed respectively with a blue and a gray square.}}
    \label{fig:gradients}
\end{figure}

Fig.~\ref{fig:gradients} shows that the ERIS iron abundances follow the well-established negative metallicity gradient of the young Galactic disk (panel~a)). The [Mg/H] and [Si/H] gradients (panels~b) and d)) are in a good agreement with the optical reference sample, with the two datasets tracing consistent trends across the explored $\mathrm{R_{GC}}$ range. 
[Mg/Fe] and [Si/Fe] (panels~c) and e)) remain remarkably flat across the entire explored range, in agreement with the findings of \citet{nunnari26} and other high-resolution studies (\citealt{genovali15, kovtyukh22, dalponte25, trentin26}), supporting a scenario in which  Mg and Si enrichment scales closely with iron in the young stellar population.

In this work, we use the chemical gradients as a consistency check for our derived Fe, Mg, and Si abundances, verifying that the observed trends match the behavior expected from the literature. The overall shape of these gradients has already been discussed in detail in our previous study (\citealt{nunnari26}, see also references therein). Here, we instead focus on the diagnostic potential of our results for probing gradient behavior in the most heavily reddened regions of the disk, namely the inner ($\mathrm{R_{GC}}\lesssim6$~kpc) and outer ($\mathrm{R_{GC}}\gtrsim 20$~kpc) disk, along with those beyond the bulge. 
Although our sample does not extend into the innermost disk, where previous studies have reported a flattening or decline in metallicity (\citealt{kovtyukh22, matsunaga23}), the present analysis constitutes a methodological step toward the tools required to investigate such features in future work. This is particularly relevant in light of recent evidence that this behavior is not unique to the Milky Way (see \citealt{sextl24, sextl25}).
In the outer disk, substantial observational effort will be needed to identify new CCs suitable for enlarging the current sample.

\section{Summary and Conclusions}
\label{sec:summary}
In this study, we performed an exploratory spectral-synthesis analysis of 27 CCs observed with ERIS@VLT in the J band at low-resolution (R~$\sim$~4000) to derive atmospheric parameters and Mg and Si abundances. To address the limited number of useful lines and blending at this resolution, we adopted a two-iteration optimization procedure in PySME with NLTE departure coefficients, which yields consistent stellar parameters and abundances and confirms the usefulness of J-band spectroscopy at low-resolution.

The derived parameters cover a wide range, with effective temperatures between approximately 4800 and 6800 K, surface gravities from 0.5 to 3.0~dex, and metallicities from about $-~0.6$ to $+~0.5$~dex, fully compatible with Galactic CCs. We find no significant trends of [Fe/H] with $\mathrm{T_{eff}}$, $\mathrm{\log g}$, or $\mathrm{v_{mic}}$, supporting the internal consistency of our method. For 12 stars, we provide the first estimates of Mg and Si abundances.

The kinematic analysis shows that almost all targets are consistent with thin-disk membership. The sample spans Galactocentric distances from roughly 5 to 28~kpc, with most stars concentrated between 5 and 10~kpc. One notable exception is Gaia DR3 5415098722470316544, located at $\mathrm{R_{GC}} \approx28$~kpc and $Z = +4.8 \pm 
1.1$~kpc, which follows a halo-like orbit and may instead be a runaway disk star. A second extreme object, ASAS~J062939-1840.5 from the comparison sample, has a large vertical displacement but remains on a disk-like orbit, hinting that the outer warp may extend farther than previously assumed.

In passing we note that in a recent investigation \citet{fiteni+26} suggested, on the basis of numerical simulations,  that the edge of the MW’s star-forming disk has break radius in the stellar density profile located between $\sim$11 and $\sim$12~kpc. This break is associated with a sharp decline in the star formation rate. Such circumstantial evidence may be linked to several mechanisms such as the dynamical influence of the bar’s outer Lindblad resonance, the onset of the Galactic warp, or thermally regulated star formation. However, the identification of two classical Cepheids with pulsation periods of $\sim$16.94 and $\sim$7.78 days, corresponding to individual ages of $\approx$30--$\approx$50~Myrs \citep{bono05}, and assuming that these stars formed close to their present position, is suggesting that star formation in the outskirts of the Galactic thin disk was not vanishing during this time interval. In addition, these two objects are located at $\mathrm{R_{GC}\approx28-29}$~kpc, in a region where star formation is expected to be extremely sparse. Both CCs are located at large Galactocentric distances and exhibit peculiar kinematic properties. However, they are still located within the fringes of neutral hydrogen distribution of the thin disk as suggested by radio measurements (\citealt{kalberla09}; \citealt{nakanishi16}). A more detailed discussion of their kinematic properties, based on the backward integration of their orbits, will be presented in a forthcoming paper (Pepe et al. in preparation).

The abundance ratios of Mg and Si are nearly constant across the sample, with mean values of $\langle [\mathrm{Mg}/\mathrm{Fe}] \rangle = 0.01\pm0.12$~dex and $\langle [\mathrm{Si}/\mathrm{Fe}] \rangle = 0.10 \pm 0.09$~dex, in excellent agreement with previous high-resolution studies. However, we cannot exclude a possible systematic of $\approx$~+0.10 dex that affects Si abundances. This results supports chemical-evolution models predicting near-solar $[\alpha/\mathrm{Fe}]$ for young thin-disk populations. The results are consistent with the expected negative iron gradient with $\mathrm{R_{GC}}$, while [Mg/Fe] and [Si/Fe] remain flat across the explored disk.

Overall, these results demonstrate that low-resolution J-band spectroscopy is a valid tool for probing the chemical structure of the Milky Way disk through CCs, especially in regions where optical observations are strongly limited by extinction. Future work with larger samples and upcoming near-infrared surveys  (MOONS, \citealt{cirasuolo12}) will further refine the disk abundance pattern and extend this approach to highly obscured regions across the Galactic bar and the nuclear stellar disk (\citealt{matsunaga11}).

\section{Data availability}
Table 3, which contains the chemo-dynamical results for the ERIS sample, is available only in electronic form at the CDS through anonymous ftp to \url{cdsarc.u-strasbg.fr} (130.79.128.5) or via \url{http://cdsweb.u-strasbg.fr/cgi-bin/qcat?J/A+A/}.


\begin{acknowledgements}
Part of the research activities described in this paper was carried out with the contribution of: i) NextGenerationEU funds within the National Recovery and Resilience Plan (PNRR), Mission 4 - Education and Research, Component 2 - From Research to Business (M4C2), Investment Line 3.1 - Strengthening and creation of Research Infrastructures, Project IR0000034 – “STILES - Strengthening the Italian Leadership in ELT and SKA” , CUP C33C22000640006; ii) BANDO PRIN 2022 - Settore ERC PE9 ‘Universe sciences’, Project 20222495WS – “Women measuring stars: a comprehensive strategy for the exploitation of adaptive optics data ” , CUP C53D23000810006. V. D'Orazi acknowledges support from the INAF Minigrant 2024 {\it MUGS}. G. Bono acknowledges partial support from the INFN InDARK project. E. Sextl and R.-P. Kudritzki acknowledge support from the Munich Excellence Cluster Origins and the Munich Institute for Astro-, Particle, and Biophysics (MIAPbP), both funded by the Deutsche Forschungsgemeinschaft (DFG, German Research Foundation) under Germany's Excellence Strategy EXC-2094 390783311. We acknowledge funding from IPRIN MUR2022 project (code 2022ARWP9C) 'Early Formation and Evolution of Bulge and Halo (EFEBHO),' PI: Marconi, M., funded by the European Union – Next Generation EU; Large Grant INAF 2023 MOVIE (P.I. M. Marconi). This research was supported by the International Space Science Institute (ISSI) in Bern/Beijing through the ISSI/ISSI-BJ International Team project ID \#24-603 – "EXPANDING Universe" (EXploiting Precision AstroNomical Distance INdicators in the Gaia Universe). J. Olivares-Carvajal acknowledges financial support from the project "Partecipazione LSST - Large Synoptic Survey Telescope" (Ob.Fu. 1.05.03.06; local PI: M. Brescia). M. Zoccali acknowledges support from ANID Fondecyt Regular 1230731. The authors thank Elena Pancino and Santi Cassisi for their valuable contribution to this work. The authors sincerely thank the anonymous referee for insightful comments and suggestions that helped us to strengthen the manuscript.
\end{acknowledgements}

\bibliographystyle{aa}
\bibliography{nunnari}

\begin{appendix}
\nolinenumbers
\section{Spectral lines blending}
\label{App:blending}
\begin{figure*}
    \centering
    \includegraphics[width=0.24\linewidth]{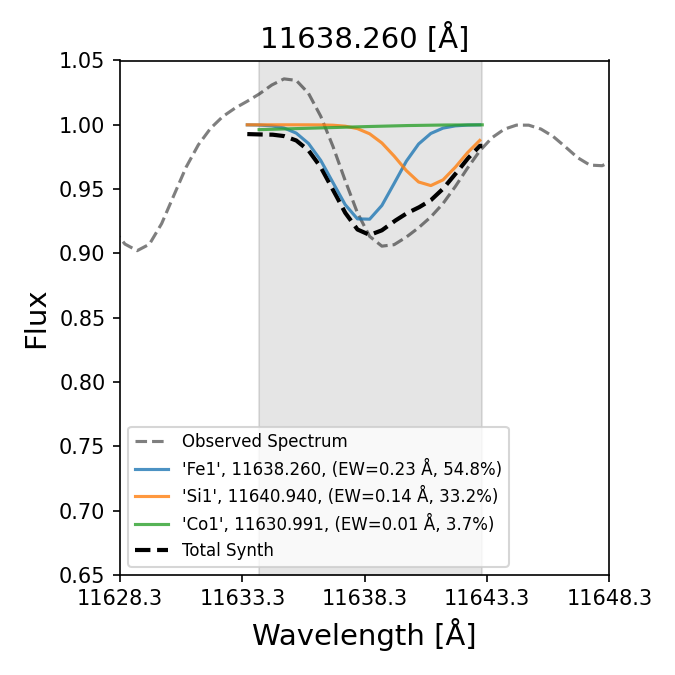}
    \includegraphics[width=0.24\linewidth]{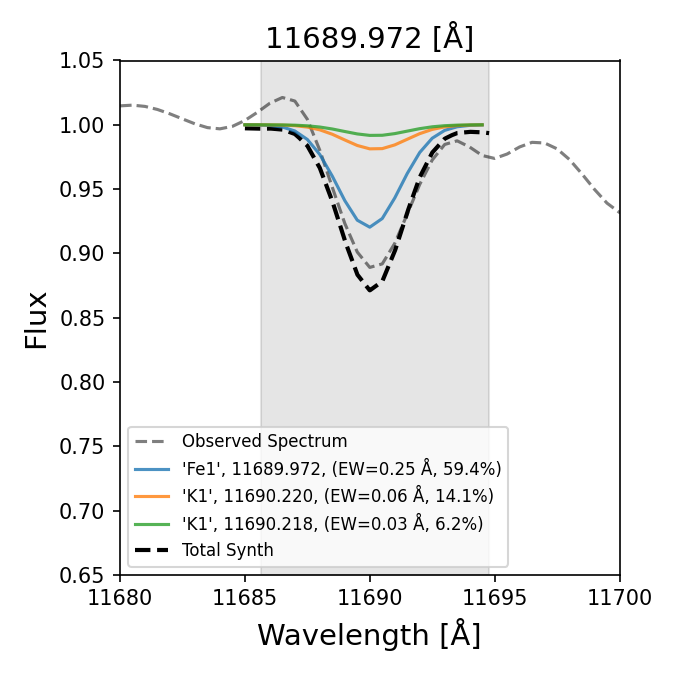}
    \includegraphics[width=0.24\linewidth]{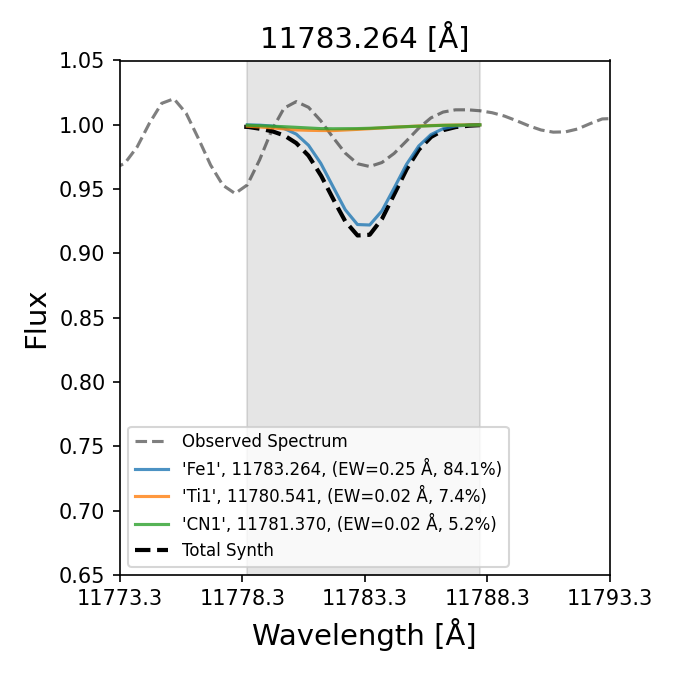}
    \includegraphics[width=0.24\linewidth]{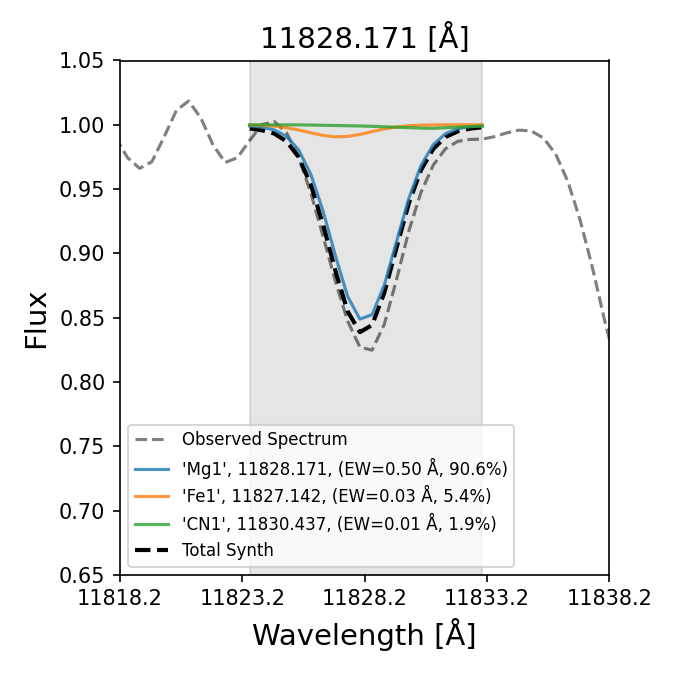}
    \includegraphics[width=0.24\linewidth]{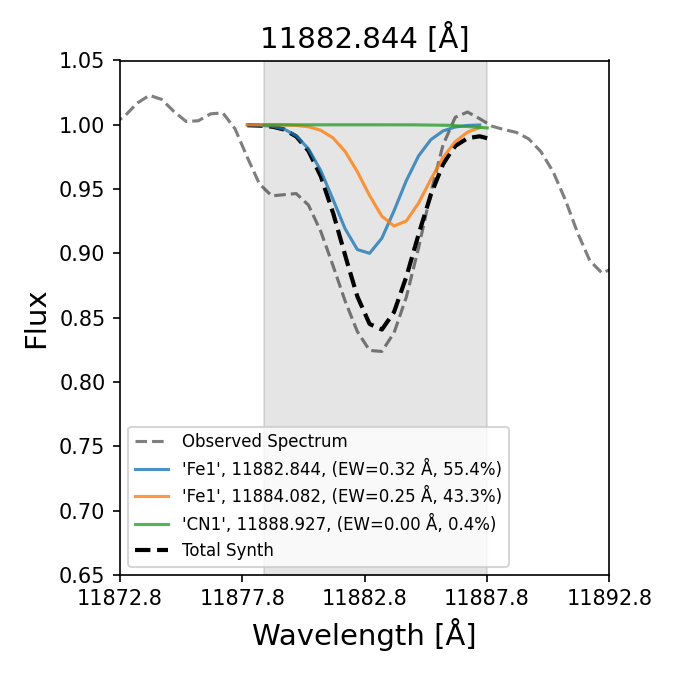}
    \includegraphics[width=0.24\linewidth]{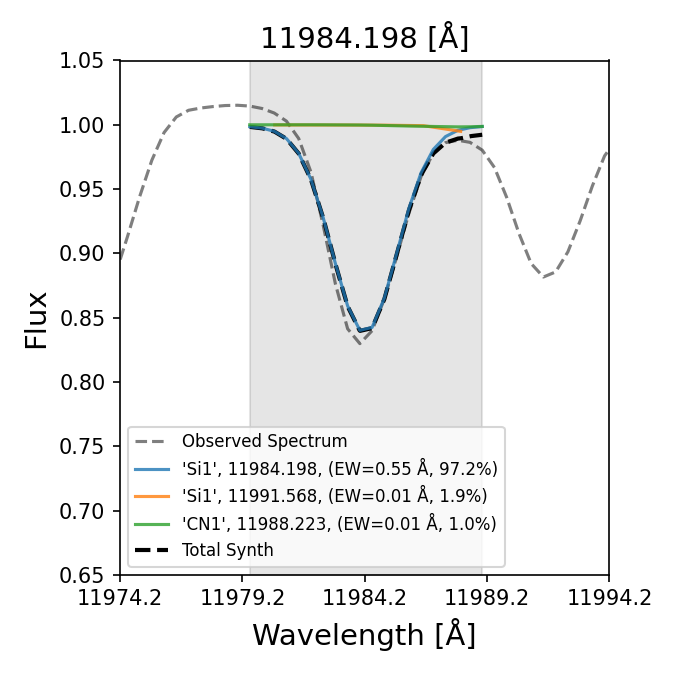}
    \includegraphics[width=0.24\linewidth]{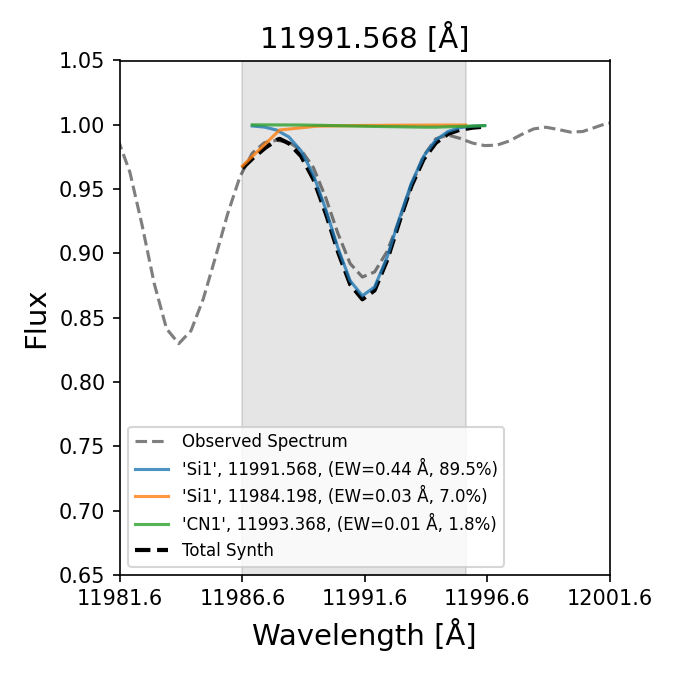}
    \includegraphics[width=0.24\linewidth]{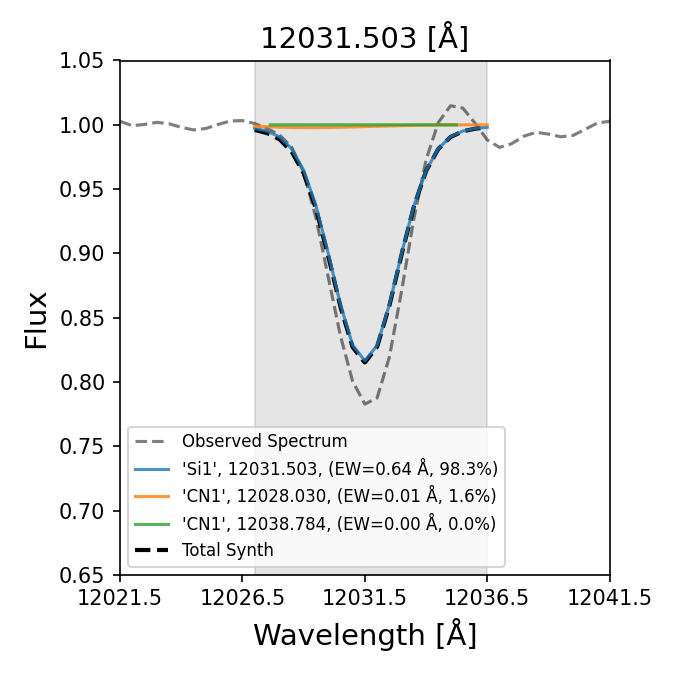}
    \includegraphics[width=0.24\linewidth]{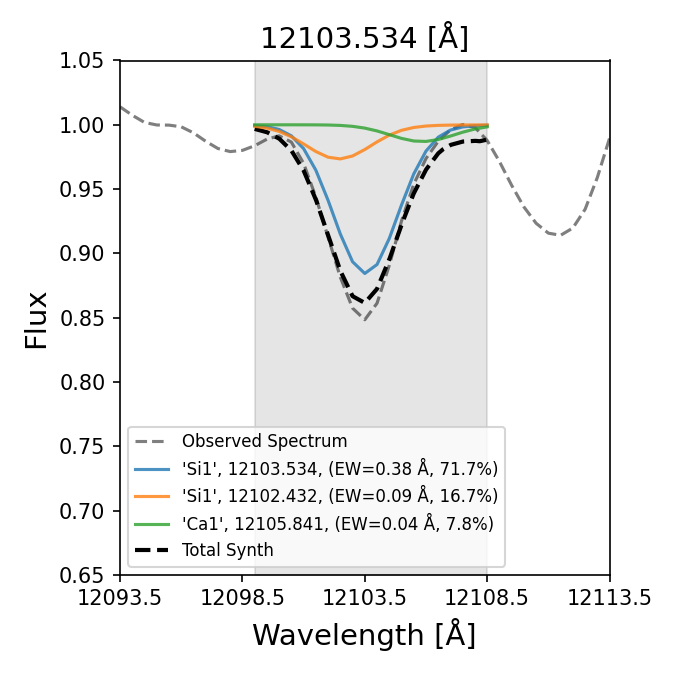}
    \includegraphics[width=0.24\linewidth]{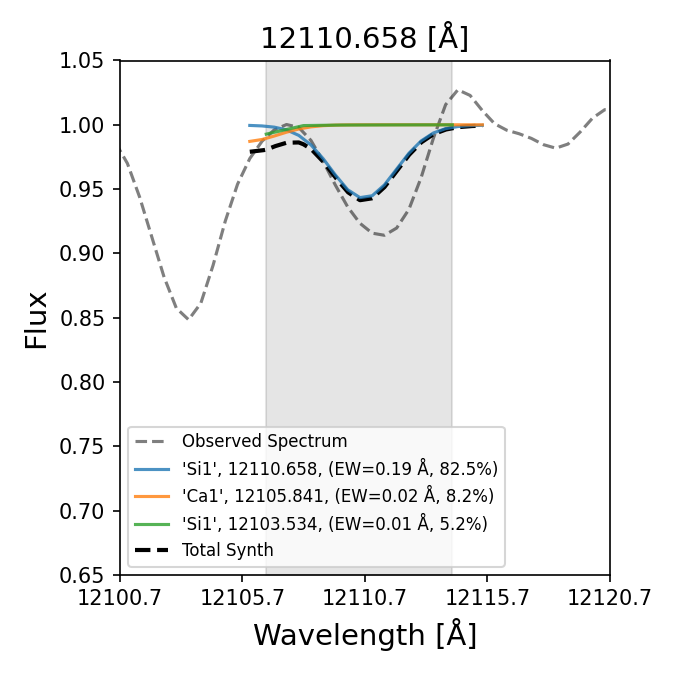}
    \includegraphics[width=0.24\linewidth]{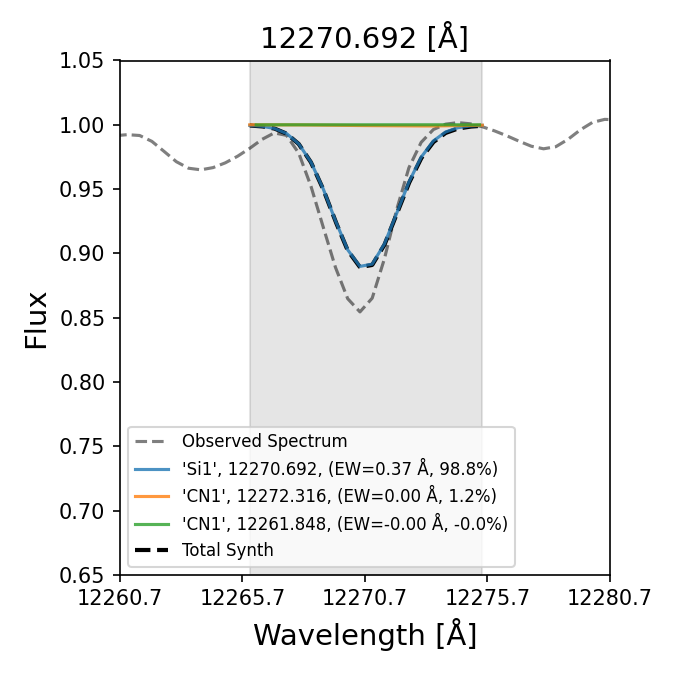}
    \includegraphics[width=0.24\linewidth]{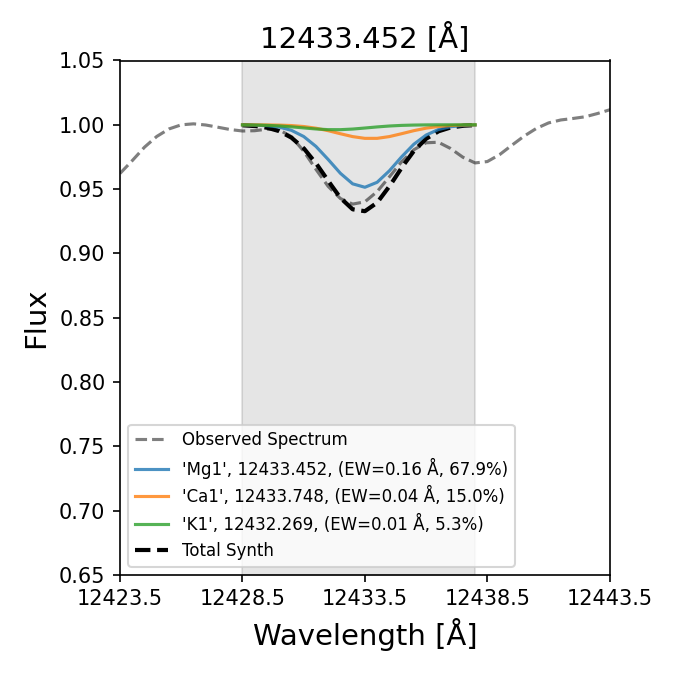}
    \caption{\small{Each panel shows the individual components of a spectral line employed in this work. In blue the primary component is drawn, while the secondary and tertiary ones are shown respectively with the orange and green lines. The legend shows the corresponding transition of the specific component, the center wavelength, the equivalenth width and percentage relative to the total profile of the line. The total synthetic spectrum is depicted with the dashed black line, while the observed spectrum of one of the CCs in our sample is shown with the dashed gray line.}}
    \label{fig:blending}
\end{figure*}

The spectra of this study have a resolution of roughly 4000, that correspond to $\Delta\lambda\approx$~0.3~nm at a typical J-band wavelength of $\lambda$~=~1200~nm. This means that two spectral lines with central wavelengths within $\approx0.3$~nm of each other are blended. We investigated the level of blending for each line employed in this study by synthesizing with PySME each line with their individual components at the atmospheric parameters of the targets, quantifying the level of the blending. In Figure \ref{fig:blending} we show an example for a relatively hot, metal-rich Cepheid, with Gaia DR3 5853901851624637824, whose determined atmospheric parameters are $\mathrm{T_{eff}=6800}$~K, $\mathrm{\log g}=2.6$~dex, $\mathrm{[Fe/H]}=0.4$~dex and $\mathrm{v_{mic}}=3.5$~$\mathrm{km~s^{-1}}$.
The blending details for each feature are the following:
\begin{itemize}
    \item $11638.26~\AA$, Fe~I: presents a strong blending with a Si~I line with central wavelength at $11640.94~\AA$.
    
    \item $11689.97~\AA$, Fe~I: affected by a blending with the hyperfine structure of K~I at around $11690.2~\AA$.

    \item $11783.26~\AA$, Fe~I: the secondary components (Ti~I and CN~I) are negligible.

    \item $11828.17~\AA$, Mg~I: is a relatively strong line (EW~$\approx$~0.50~$\AA$) and the secondary components (Fe~I and CN~I) contribute less than 10\%.

    \item $11882.84~\AA$, Fe~I: is blending with another Fe~I line with central wavelength at 11884.08~$\AA$ creating a global line profile dominated by Fe~I transitions.

    \item $11984.20~\AA$, Si~I: is a strong line (EW~$\approx$~0.55~$\AA$) with secondary components below the 5\%.

    \item $11991.57~\AA$, Si~I: same as the previous line, but with EW$\approx0.44~\AA$ and with a secondary component below the 9\%.

    \item $12031.50~\AA$, Si~I: same as the previous line, but with even higher EW ($\approx0.60~\AA$).

    \item $12103.53~\AA$, Si~I: presents significant secondary contributions, that are the Si~I at $12102.43~\AA$ ($\approx~17~\%$) and Ca~I at $12105.84~\AA$ ($\approx8~\%$).

    \item $12110.66~\AA$, Si~I: presents a weak blending on the left wing with Ca~I at $12105.84~\AA$ and Si~I at $12103.53~\AA$.

    \item $12270.69~\AA$, Si~I: the secondary components are negligible.




    \item $12433.45 \, \AA$, Mg I: presents a significant secondary (Ca I at $12433.75\,\AA$, $\approx15\%$) and tertiary component (K I at $12432.27\,\AA$, $\approx5\%$).

\end{itemize}

\begin{figure}[!]
    \centering
    \includegraphics[width=1.\linewidth]{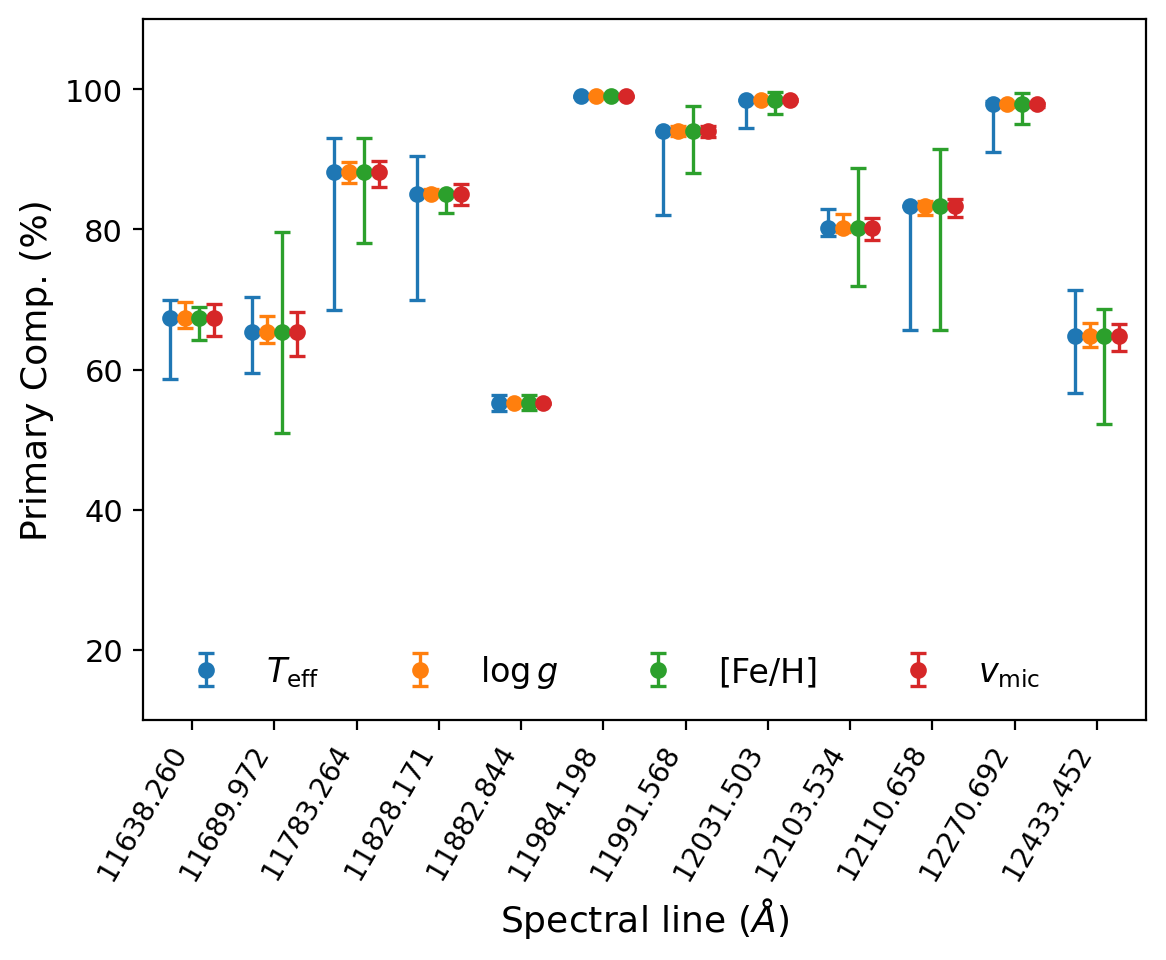}
    \caption{\small{Percentage contribution of the primary component to the total line profile for each spectral line. Each point represents the percentage obtained using the reference stellar parameters, while the vertical bars indicate the deviations from the reference value resulting from variations in $\mathrm{T_{eff}}$ (blue),  $\mathrm{\log g}$ (orange), $\mathrm{[Fe/H]}$ (green), and $\mathrm{v_{mic}}$ (red).}}
    \label{percentage}
\end{figure}

To quantify the sensitivity of the blending level of the primary component to variations in the stellar parameters, we computed its percentage contribution to the total line profile using a reference set of parameters: 5500~K, 1.5~dex, 0~dex, and 4~$\mathrm{km~s^{-1}}$, respectively for $\mathrm{T_{eff}}$, $\mathrm{\log g}$, $\mathrm{[Fe/H]}$, and $\mathrm{v_{mic}}$. We then varied each parameter independently; the temperature by 500~K, the gravity by 1~dex, the metallicity by 0.5~dex and the microturbulence by 1~$\mathrm{km~s^{-1}}$ with respect to the reference values. Figure~\ref{percentage} shows the resulting changes in the blending percentages relative to the reference model for each spectral line.

\section{Metallicity comparison with Gaia-based and photometric studies}\label{12ccs}
In this section, we compare the metallicities derived for our sample of 27 CCs with results from five recent studies that provided estimates for the same objects. These studies employ methodological approaches that differ fundamentally from each other and from the physical spectral synthesis applied in this work.

\citet{hocde23} uses a purely photometric approach. The authors calibrated empirical relations linking metallicity to the morphology of the light curve in the V and I bands. \citet{khalatyan24} utilizes low-resolution Gaia XP spectra (R~$\approx$~40--150). The method employs the XGBoost gradient-boosted random-forest regressor to transfer stellar "labels" (atmospheric parameters) from high-resolution ground-based surveys to hundreds of millions of Gaia sources. \citet{huson25} uses Gaia XP spectra, adopting a convolutional neural network called Gaia Net. Since the Gaia XP spectra are the result of co-adding many transits over time, the derived parameters represent a temporal average of the Cepheid pulsation cycle. \citet{ardern-arentsen25} also relies on Gaia XP spectra and an artificial neural network. While primarily focused on metal-poor and carbon-enhanced stars, the study provides atmospheric parameters for a vast sample of Gaia giants, including several Cepheids in our sample. \citet{matsuno24} adopts a hybrid approach using Gaia RVS spectra (R~$\approx$~11500) centered on the Calcium triplet. The authors combined equivalent width measurements with photometric effective temperatures and surface gravities (derived from Gaia and 2MASS) to break parameter degeneracies.

\begin{figure}[h]
    \centering
    \includegraphics[width=1.\linewidth]{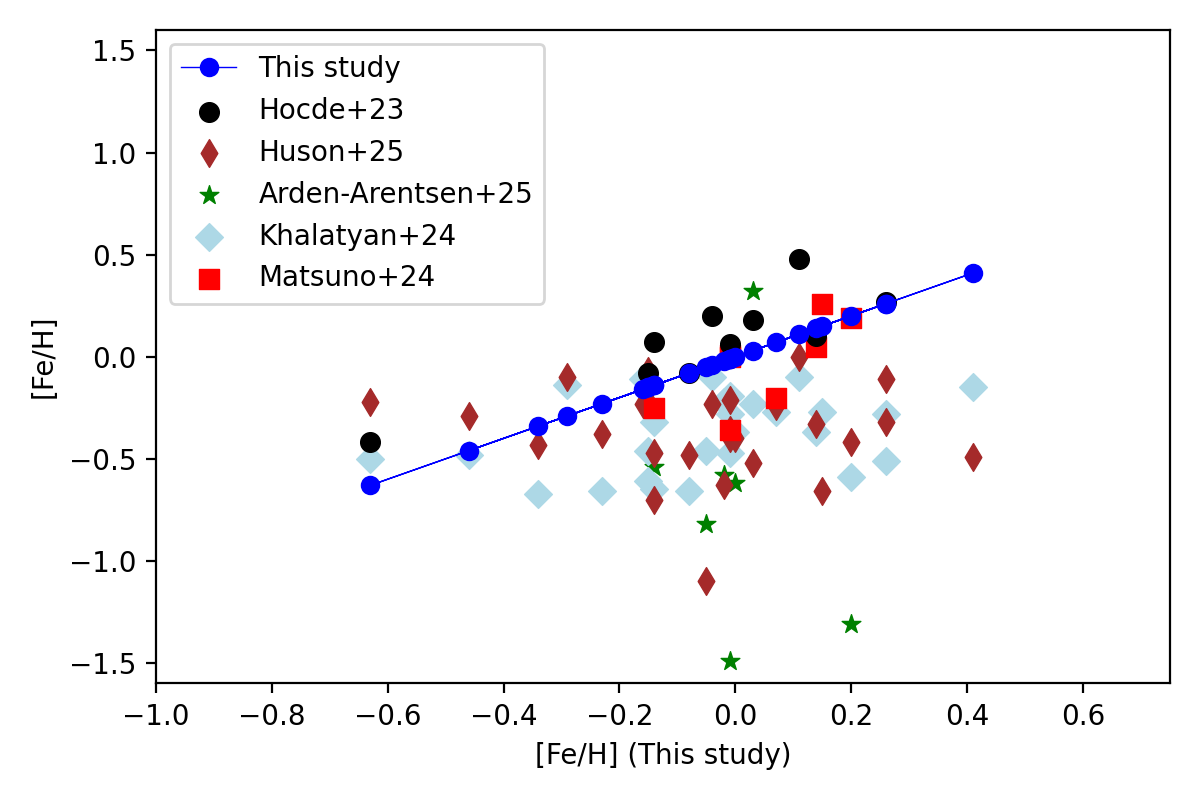}
    \caption{\small{Comparison of the [Fe/H] derived in this study from ERIS J-band low-resolution spectroscopy with estimates reported in the literature for the same targets. The solid blue line represents the 1-to-1 correspondence. Reference studies are indicated by different symbols: black circles for the photometric estimates of \citet{hocde23}, brown diamonds for the Gaia Net results of \citet{huson25}, green stars for the neural network predictions of \citet{ardern-arentsen25}, light blue diamonds for the SHBoost values of \citet{khalatyan24}, and red squares for the RVS-based estimates of \citet{matsuno24}.
}}
    \label{fig:comp_gaia}
\end{figure}

Figure~\ref{fig:comp_gaia} illustrates the comparison between the [Fe/H] values derived in this study (ERIS J-band) and those reported in the five aforementioned works. The blue line represents the 1-to-1 correspondence. Significant scatter is observed, particularly for the results from \citet[brown diamonds]{huson25}, \citet[light blue diamonds]{khalatyan24} and \citet[green stars]{ardern-arentsen25}. Data from \citet[red squares]{matsuno24} and \citet[black circles]{hocde23} show a better correlation with our findings. The use of RVS or photometric data appears to yield results more consistent with our metallicity scale. It is crucial to emphasize that while the literature studies are predominantly "data-driven" or statistical, our approach utilizes case-by-case spectral synthesis. Furthermore, our J-band spectroscopy mitigates the impact of extinction—a factor that can introduce significant bias in methods based on optical photometry or XP spectra.

\section{Atmospheric parameters comparison}\label{App:atm}
\begin{figure}
    \centering
    \includegraphics[width=1.\linewidth]{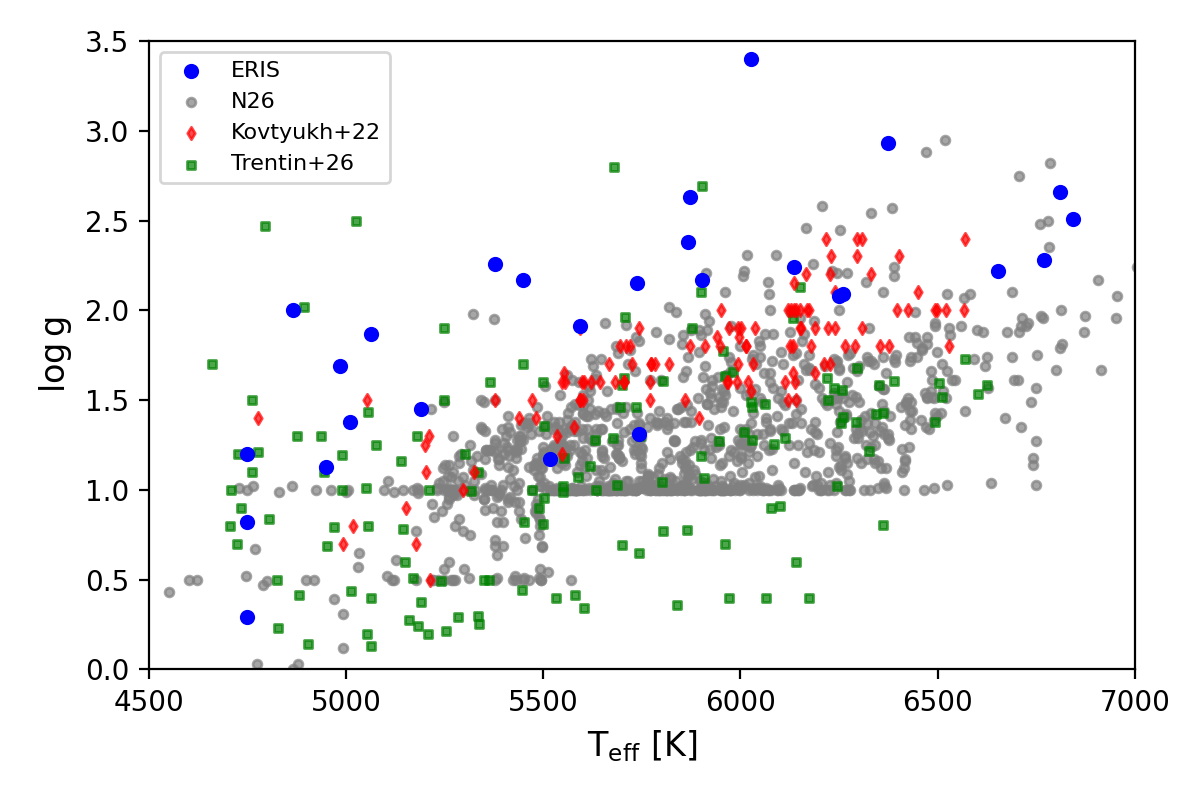}
    \caption{\small{Distribution of the atmospheric parameters derived in this study (ERIS, blue circles) on the $\mathrm{\log g-T_{eff}}$ plane, compared with those from \citet[gray points]{nunnari26}, \citet[red diamonds]{kovtyukh22}, and \citet[green squares]{trentin26}.}}
    \label{fig:t_g}
\end{figure}

Figure~\ref{fig:t_g} shows the distribution of the ERIS atmospheric parameters on the $\mathrm{\log g-T_{eff}}$ plane, alongside three literature samples of CCs. The ERIS targets broadly follow the same locus defined by the comparison samples, confirming that the derived parameters are physically consistent with those expected for CCs. However, the ERIS points tend to be systematically displaced towards higher $\mathrm{\log g}$ values at comparable $\mathrm{T_{eff}}$. This offset may be due to the intrinsically larger uncertainties on $\mathrm{\log g}$ derived from low-resolution J-band spectra, where the sensitivity of the employed line list to surface gravity is reduced compared to the optical line lists.

\begin{figure}
    \centering
    \includegraphics[width=1.\linewidth]{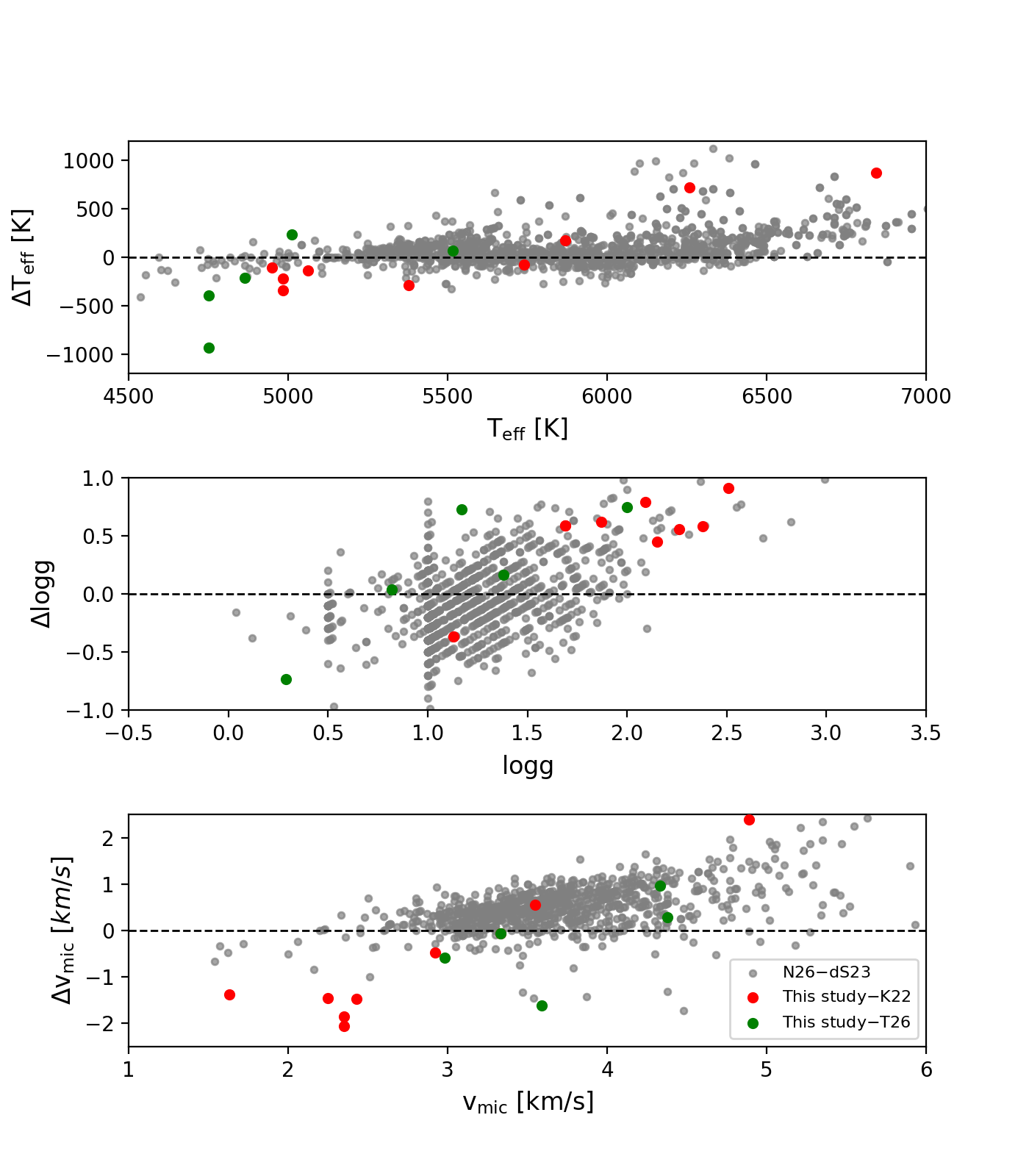}
    \caption{\small{Residuals of the $\mathrm{T_{eff}}$ (top panel), $\mathrm{\log g}$ (central panel) and $\mathrm{v_{mic}}$} (bottom panel) between this study and the literature. Red points show the differences with respect to \citet[K22]{kovtyukh22}, green points those with respect to \citet[T26]{trentin26}, and gray points the residuals from \citet[N26]{nunnari26} with respect to \citet[dS23]{da2023oxygen}, shown as a reference for the typical scatter expected from independent analyses of CCs.}
    \label{fig:delta}
\end{figure}

We also measured the difference in atmospheric parameters for the same CCs between different studies. The residuals shown in Fig.~\ref{fig:delta} exhibit a non-negligible scatter around zero for all three atmospheric parameters, with deviations reaching up to $\approx$~1000~K in $\mathrm{T_{eff}}$, $\approx$~1~dex in $\mathrm{\log g}$, and $\approx~\mathrm{2~km~s^{-1}}$ in $\mathrm{v_{mic}}$. The measured difference is the sum of a method-dependent and a pulsational-phase difference. A significant contribution to this scatter is expected to arise from the pulsational nature of CCs: since the atmospheric parameters vary continuously along the pulsation cycle, any two spectra of the same star obtained at different pulsational phases will probe intrinsically different atmospheric conditions (\citealt{da2022new}). The gray points in Fig.~\ref{fig:delta}, representing the comparison between two independent optical analyzes of a much larger sample (\citealt{da2023oxygen, nunnari26}), confirm that residuals of this magnitude are consistent with those expected even between high-resolution optical studies of CCs. 

Finally, we computed the pulsation phases of the ERIS and literature spectra and found only two stars with a phase difference of 0.03 or less:
\begin{itemize}
    \item Gaia DR3 5862711040048310528 (\citealt{trentin26}; phase difference 0.03). $\mathrm{T_{eff}}=4865\pm112$~K against $5076\pm95$~K (1.4~$\sigma$); $\mathrm{\log g}=2.0\pm0.5$~dex against $1.25\pm0.05$~dex (consistent within 1~$\sigma$); $\mathrm{v_{mic}}=3.3\pm0.3~\mathrm{km~s^{-1}}$ against $3.39\pm0.44~\mathrm{km~s^{-1}}$ (fully consistent);
    \item Gaia DR3 5409250351407633536 (\citealt{kovtyukh22}; phase difference 0.01). $\mathrm{T_{eff}}=5377\pm195$~K against $5668\pm93$~K (1.3~$\sigma$); $\mathrm{\log g}=2.26\pm0.4$~dex against $1.7\pm0.2$~dex (1.3$~\sigma$); $\mathrm{v_{mic}}=1.63\pm0.27~\mathrm{km~s^{-1}}$ against $3.0\pm0.3~\mathrm{km~s^{-1}}$ (3.4~$\sigma$).
\end{itemize}
It is worth mentioning that in \citet{nunnari26} we showed that our NLTE approach yields values lower by about 0.15~$\mathrm{km~s^{-1}}$ on average, and up to 0.6$~\mathrm{km~s^{-1}}$ for individual stars; this accounts for only part of the difference.

\section{Light curves of Gaia DR3 5415098722470316544 and ASAS J062939-1840.5}\label{app:light_curve}

\begin{figure}
    \centering
    \includegraphics[width=0.9\linewidth]{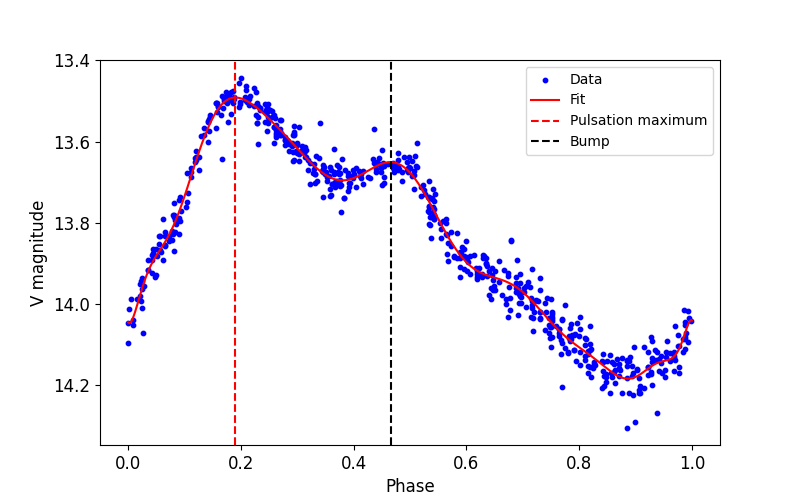}
    \includegraphics[width=0.9\linewidth]{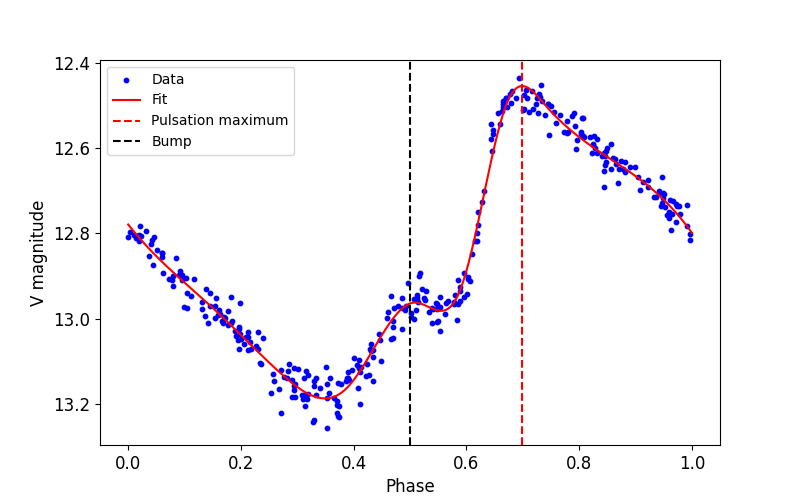}
    \caption{\small{V-band light curves of Gaia DR3 ID 5415098722470316544 (top panel, period of 7.78 days) and of ASAS J062939-1840.5 (bottom panel, period of 16.94 days). Blue points are the observed V mag from ASAS, the red line is a polynomial fit, the red and the black dashed lines identify respectively the pulsation maximum and the bump.}}
    \label{fig:lightcurve}
\end{figure}

We used the ASAS-SN Variable Stars Database\footnote{\url{https://asas-sn.osu.edu/variables/}} (\citealt{shappee14,kochanek17,hart23}) to verify the classification of Gaia DR3 ID 5415098722470316544 and ASAS~J062939-1840.5 as CCs. Figure~\ref{fig:lightcurve} displays the V-band light curves, whose morphology is consistent with a CC. In particular, the light curve of Gaia DR3 ID 5415098722470316544 exhibits a characteristic bump, a feature known as the "Hertzsprung progression" (\citealt{marconi24}), which is observed on the descending branch of Classical Cepheid light curves with periods in the range $\approx 6-9$ days. For such periods, the bump occurs after the pulsation maximum in phase and at a lower brightness level. As shown in the top panel of Fig~\ref{fig:lightcurve}, the bump is located at a phase later than the pulsation maximum, consistent with the expected behavior for a Galactic CC with a period of 7.78 days, since the bump progressively shifts toward the pulsation maximum as the period increases, eventually merging with it at periods of 9.8$\pm$0.2 days (\citealt{bono2024cepheids, marconi24}). ASAS~J062939-1840.5 has a larger period and its light curve shows the bump in the rising branch. 

\end{appendix}
\end{document}